\pdfoutput=1

\documentclass[12pt,a4paper]{article}
\newcommand{\unitm}[1]{\,{\rm{#1}}}
\renewcommand*{\thefootnote}{\fnsymbol{footnote}}
\usepackage{ifthen} 
\newboolean{pdflatex}
\setboolean{pdflatex}{true} 
\newboolean{articletitles}
\setboolean{articletitles}{true} 

\newboolean{uprightparticles}
\setboolean{uprightparticles}{false} 

\newboolean{inbibliography}
\setboolean{inbibliography}{false} 

\usepackage{microtype}
\usepackage{lineno}  
\usepackage{xspace} 
\usepackage{caption} 

\usepackage{graphicx}  
\usepackage{color}
\usepackage{colortbl}
\graphicspath{{./figs/}} 

\usepackage{amsmath} 
\usepackage{amssymb}
\usepackage{amsfonts}
\usepackage{upgreek} 

\newcommand*\patchAmsMathEnvironmentForLineno[1]{%
\expandafter\let\csname old#1\expandafter\endcsname\csname #1\endcsname
\expandafter\let\csname oldend#1\expandafter\endcsname\csname
end#1\endcsname
 \renewenvironment{#1}%
   {\linenomath\csname old#1\endcsname}%
   {\csname oldend#1\endcsname\endlinenomath}%
}
\newcommand*\patchBothAmsMathEnvironmentsForLineno[1]{%
  \patchAmsMathEnvironmentForLineno{#1}%
  \patchAmsMathEnvironmentForLineno{#1*}%
}
\AtBeginDocument{%
\patchBothAmsMathEnvironmentsForLineno{equation}%
\patchBothAmsMathEnvironmentsForLineno{align}%
\patchBothAmsMathEnvironmentsForLineno{flalign}%
\patchBothAmsMathEnvironmentsForLineno{alignat}%
\patchBothAmsMathEnvironmentsForLineno{gather}%
\patchBothAmsMathEnvironmentsForLineno{multline}%
\patchBothAmsMathEnvironmentsForLineno{eqnarray}%
}

\usepackage{hyperref}    
\usepackage[all]{hypcap} 

\def\lhcb {\mbox{LHCb}\xspace}

\def\MagUp {\mbox{\em Mag\kern -0.05em Up}\xspace}

\ifthenelse{\boolean{uprightparticles}}%
{

 \def\PDelta      {\ensuremath{\Delta}\xspace}                 
 \def\PXi      {\ensuremath{\Xi}\xspace}                 
 \def\PLambda      {\ensuremath{\Lambda}\xspace}                 
 \def\PSigma      {\ensuremath{\Sigma}\xspace}                 
 \def\POmega      {\ensuremath{\Omega}\xspace}                 
 \def\PUpsilon      {\ensuremath{\Upsilon}\xspace}                 
 
 \def\PB      {\ensuremath{\mathrm{B}}\xspace}                 
                  
 \def\PD      {\ensuremath{\mathrm{D}}\xspace}

 \def\PK      {\ensuremath{\mathrm{K}}\xspace}

 \def\Pi      {\ensuremath{\mathrm{i}}\xspace}

}
{

 \mathchardef\PDelta="7101
 \mathchardef\PXi="7104
 \mathchardef\PLambda="7103
 \mathchardef\PSigma="7106
 \mathchardef\POmega="710A
 \mathchardef\PUpsilon="7107
                  
 \def\PB      {\ensuremath{B}\xspace}                 
                  
 \def\PD      {\ensuremath{D}\xspace}

 \def\PK      {\ensuremath{K}\xspace}

 \def\Pi      {\ensuremath{i}\xspace}

}

\makeatletter
\ifcase \@ptsize \relax
  \newcommand{\miniscule}{\@setfontsize\miniscule{4}{5}}
\or
  \newcommand{\miniscule}{\@setfontsize\miniscule{5}{6}}
\or
  \newcommand{\miniscule}{\@setfontsize\miniscule{5}{6}}
\fi
\makeatother

\DeclareRobustCommand{\optbar}[1]{\shortstack{{\miniscule (\rule[.5ex]{1.25em}{.18mm})}
  \\ [-.7ex] $#1$}}

  \def\Kbar    {{\kern 0.2em\overline{\kern -0.2em \PK}{}}\xspace}

\def\KorKbar    {\kern 0.18em\optbar{\kern -0.18em K}{}\xspace}

  \def\Dbar    {{\kern 0.2em\overline{\kern -0.2em \PD}{}}\xspace}

\def\DorDbar    {\kern 0.18em\optbar{\kern -0.18em D}{}\xspace}

\def\Bbar    {{\ensuremath{\kern 0.18em\overline{\kern -0.18em \PB}{}}}\xspace}

\def\BorBbar    {\kern 0.18em\optbar{\kern -0.18em B}{}\xspace}

  \def\Y#1S{\ensuremath{\PUpsilon{(#1S)}}\xspace}

\def\Lbar        {{\ensuremath{\kern 0.1em\overline{\kern -0.1em\PLambda}}}\xspace}
\def\LorLbar    {\kern 0.18em\optbar{\kern -0.18em \PLambda}{}\xspace}

\def\AT#1     {\ensuremath{A_{\mathrm{T}}^{#1}}\xspace}           

\def\C#1      {\ensuremath{\mathcal{C}_{#1}}\xspace}                       
\def\Cp#1     {\ensuremath{\mathcal{C}_{#1}^{'}}\xspace}                    
\def\Ceff#1   {\ensuremath{\mathcal{C}_{#1}^{\mathrm{(eff)}}}\xspace}        
\def\Cpeff#1  {\ensuremath{\mathcal{C}_{#1}^{'\mathrm{(eff)}}}\xspace}       
\def\Ope#1    {\ensuremath{\mathcal{O}_{#1}}\xspace}                       
\def\Opep#1   {\ensuremath{\mathcal{O}_{#1}^{'}}\xspace}                    

\newcommand{\tev}{\ifthenelse{\boolean{inbibliography}}{\ensuremath{~T\kern -0.05em eV}\xspace}{\ensuremath{\mathrm{\,Te\kern -0.1em V}}}\xspace}
\newcommand{\gev}{\ensuremath{\mathrm{\,Ge\kern -0.1em V}}\xspace}
\newcommand{\mev}{\ensuremath{\mathrm{\,Me\kern -0.1em V}}\xspace}
\newcommand{\kev}{\ensuremath{\mathrm{\,ke\kern -0.1em V}}\xspace}
\newcommand{\ev}{\ensuremath{\mathrm{\,e\kern -0.1em V}}\xspace}
\newcommand{\gevc}{\ensuremath{{\mathrm{\,Ge\kern -0.1em V\!/}c}}\xspace}
\newcommand{\mevc}{\ensuremath{{\mathrm{\,Me\kern -0.1em V\!/}c}}\xspace}
\newcommand{\gevcc}{\ensuremath{{\mathrm{\,Ge\kern -0.1em V\!/}c^2}}\xspace}
\newcommand{\gevgevcccc}{\ensuremath{{\mathrm{\,Ge\kern -0.1em V^2\!/}c^4}}\xspace}
\newcommand{\mevcc}{\ensuremath{{\mathrm{\,Me\kern -0.1em V\!/}c^2}}\xspace}

\def\gsim{{~\raise.15em\hbox{$>$}\kern-.85em
          \lower.35em\hbox{$\sim$}~}\xspace}
\def\lsim{{~\raise.15em\hbox{$<$}\kern-.85em
          \lower.35em\hbox{$\sim$}~}\xspace}

\def\geant      {\mbox{\textsc{Geant4}}\xspace}

\def\tell1  {TELL1\xspace}
\def\ukl1   {UKL1\xspace}

\usepackage{cite} 
\usepackage{mciteplus}

\usepackage{longtable} 

\begin{document}

\renewcommand{\thefootnote}{\fnsymbol{footnote}}
\setcounter{footnote}{1}


\begin{titlepage}
\pagenumbering{roman}

\vspace*{-1.5cm}
\vspace*{1.5cm}
\hspace*{-0.5cm}
\begin{tabular*}{\linewidth}{lc@{\extracolsep{\fill}}r}
\ifthenelse{\boolean{pdflatex}}
{\vspace*{-2.7cm}\mbox{\!\!\!\includegraphics[width=.14\textwidth]{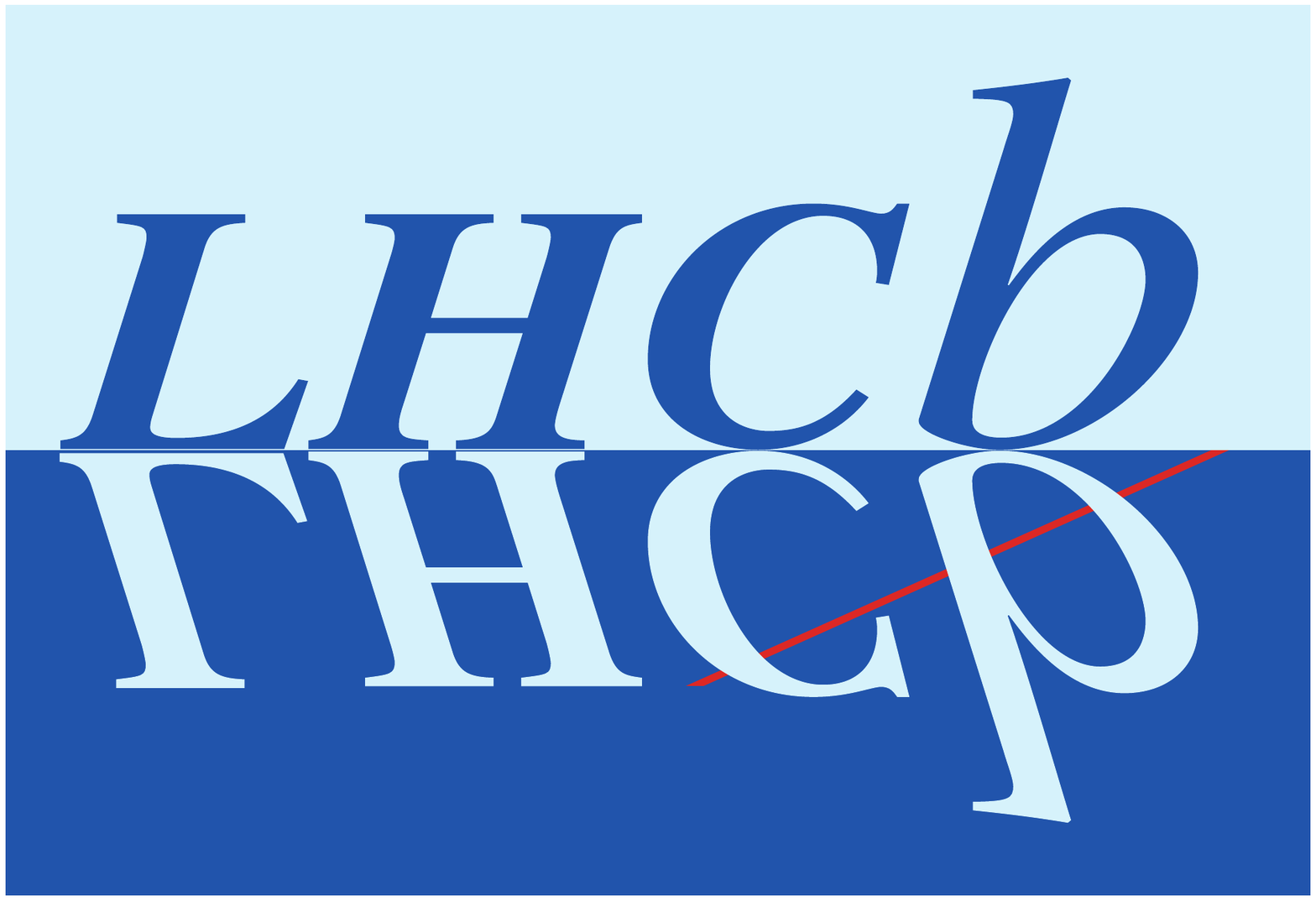}} & &}%
{\vspace*{-1.2cm}\mbox{\!\!\!\includegraphics[width=.12\textwidth]{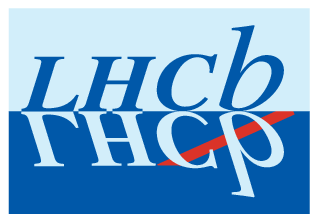}} & &}%
\\
 & & LHCb-PUB-2016-019 \\  
& & August 4, 2016 \\ 
 & & \\
\end{tabular*}

\vspace*{1.0cm}

{\bf\boldmath\huge
\begin{center}
Test of the photon detection system for the LHCb RICH 
  Upgrade in a charged particle beam
\end{center}
}

\vspace*{0.5cm}

\begin{center}
M.~K.~Baszczyk$^{5,d}$, M.~Benettoni$^4$, R.~Calabrese$^{1,a}$,
R.~Cardinale$^{2,b,}\footnote[1]{Corresponding author}$, P.~Carniti$^{3, c}$, L.~Cassina$^{3, c}$,
G.~Cavallero$^{2,b}$, L.~Cojocariu$^{f}$,  A.~Cotta Ramusino$^{1}$, C.~D'Ambrosio$^{7}$, P.~A.~Dorosz$^{5,d}$, S.~Easo$^{9}$, S.~Eisenhardt$^{10}$, M.~Fiorini$^{1,a}$, C.~Frei$^{7}$,
S.~Gambetta$^{2,b,\dagger}$, V.~Gibson$^{8}$, C.~Gotti$^{3, c}$, N.~Harnew$^{13}$, J.~He$^{7,\ddagger}$, F.~Keizer$^{8}$, W.~Kucewicz$^{5,d}$,
F.~Maciuc$^{6}$, 
M.~Maino$^{3, c}$, R.~Malaguti$^{1}$, C.~Matteuzzi$^{3}$,
M.~McCann$^{12}$, A.~Morris$^{10}$, F.~Muheim$^{10}$,
A.~Papanestis$^{9}$, G.~Pessina$^{3}$, A.~Petrolini$^{2,b}$, D.~Piedigrossi$^{7}$,
A.~Pistone$^{2,b}$, V.M.~Placinta$^{6}$, S.~Sigurdsson$^{8}$, G.~Simi$^{4,e}$, J.~Smith$^{8}$,
P.~Spradlin$^{11}$, L.~Tomassetti$^{1,a}$, S.~A.~Wotton$^{8}$. 
\bigskip\\
{\it\footnotesize
$ ^{1}$Sezione INFN di Ferrara, Ferrara, Italy\\
$^{2}$Sezione INFN di Genova, Genova, Italy\\
$ ^{3}$Sezione INFN di Milano Bicocca, Milano, Italy\\
$ ^{4}$Sezione INFN di Padova, Padova, Italy\\
$ ^{5}$Henryk Niewodniczanski Institute of Nuclear Physics  Polish Academy of Sciences, Krak{\'o}w, Poland\\
$ ^{6}$Horia Hulubei National Institute of Physics and Nuclear
Engineering, Bucharest-Magurele, Romania\\
$ ^{7}$European Organization for Nuclear Research (CERN), Geneva, Switzerland\\
$ ^{8}$Cavendish Laboratory, University of Cambridge, Cambridge, United Kingdom\\
$ ^{9}$STFC Rutherford Appleton Laboratory, Didcot, United Kingdom\\
$ ^{10}$School of Physics and Astronomy, University of Edinburgh, Edinburgh, United Kingdom\\
$ ^{11}$School of Physics and Astronomy, University of Glasgow, Glasgow, United Kingdom\\
$ ^{12}$Imperial College London, London, United Kingdom\\
$ ^{13}$Department of Physics, University of Oxford, Oxford, United
Kingdom\\
$ ^{a}$Universit{\`a} di Ferrara, Ferrara, Italy\\
$ ^{b}$Universit{\`a} di Genova, Genova, Italy\\
$ ^{c}$Universit{\`a} di Milano Bicocca, Milano, Italy\\
$ ^{d}$AGH - University of Science and Technology, Faculty of Computer Science, Electronics and Telecommunications, Krak{\'o}w, Poland\\
$ ^{e}$Universit{\`a} di Padova, Padova, Italy\\
$ ^{f}$Stefan cel Mare University of Suceava, Romania\\
$\dagger$ Now at $^{10}$\\
$\ddagger$ Now at University of Chinese Academy of Sciences, Beijing,
China
}
\end{center}
\vskip 0.5cm
\vspace{\fill}

\begin{abstract}
  \noindent
 The LHCb detector will be upgraded to make more efficient use of the available luminosity
  at the LHC in Run III and extend its potential for discovery. The
  Ring Imaging Cherenkov detectors are key
  components of the LHCb detector for particle identification. 
  In this paper we describe the setup and the results of tests in a charged particle beam, carried out to assess prototypes of 
  the upgraded opto-electronic chain
  from the Multi-Anode PMT 
  photosensor to the readout and data acquisition system.  
  
\end{abstract}

\vspace*{2.0cm}

\begin{center}
 Submitted to JINST 
\end{center}

\vspace{\fill}

{\footnotesize 
\centerline{\copyright~CERN on behalf of the \lhcb collaboration, licence \href{http://creativecommons.org/licenses/by/4.0/}{CC-BY-4.0}.}}
\vspace*{2mm}

\end{titlepage}


\newpage
\setcounter{page}{2}
\mbox{~}
%
%
%
%

\cleardoublepage


\renewcommand{\thefootnote}{\arabic{footnote}}
\setcounter{footnote}{0}

\tableofcontents


\pagestyle{plain} 
\setcounter{page}{1}
\pagenumbering{arabic}


%
\clearpage
\section{Introduction}
\label{sec:Introduction}
The LHCb experiment~\cite{Alves:2008zz} performs high-precision measurements of
CP violation and searches for New Physics, taking advantage of the 
significantly enhanced production of beauty and
charm hadrons at the Large Hadron Collider (LHC) at CERN. 
Ring Imaging Cherenkov (RICH) detectors are
fundamental to the particle identification system of the experiment
and are essential for most of the physics results published by LHCb.
The two RICH detectors have performed very
successfully during Run 1 of the LHC~\cite{LHCb-DP-2012-003}.

In order to expand the potential for discovery and the study of new phenomena 
at the LHCb experiment, an upgrade of the detector is planned for Run III.
A principal feature of the upgrade is to read out the detector at
every bunch crossing, a rate of 40~MHz, and apply a more flexible software-based trigger
system to improve the selection of interesting events at a luminosity of 
$2\times10^{33}$cm$^{-2}$s$^{-1}$.
As a consequence, the current RICH photon detectors (Hybrid Photon Detectors), with encapsulated front end electronics, will have to be replaced.
Multi-anode Photo-Multiplier Tubes (MaPMT), offering similar pixel
size, are candidates for the replacement. A detailed description of the RICH upgrade project is
given in the Technical Design Report~\cite{LHCb-TDR-014}.

Tests of the full opto-electronic chain have been performed in a
charged particle beam during Autumn 2014 at the SPS facility at CERN.  The performance of the
proposed photon detectors (Hamamatsu R11265 MaPMT~\cite{Cadamuro:2014hza})
and the feasibility
of the readout and DAQ chain, which includes
an external front-end custom readout chip (CLARO~\cite{Carniti:2012ue})
and the associated data acquisition electronics (Digital Boards), were assessed.

\section{Experimental setup}
\label{sec:expSetup}
The beam tests were performed in the North Area of the Prevessin site at CERN. A
beam consisting mainly of pions and protons with momentum of 180~\gevc was
obtained from the SPS facility and guided through a light-tight box
containing a glass planoconvex lens, used as the Cherenkov
radiator. The Cherenkov light was detected with MaPMTs housed in an aluminium
structure attached to the box. The box was placed downstream of a tracking 
telescope able to record and reconstruct the trajectory of the
incoming beam particles.

\subsection{Optical system}
\begin{figure}[htbp]
\centering
\includegraphics[scale=0.575]{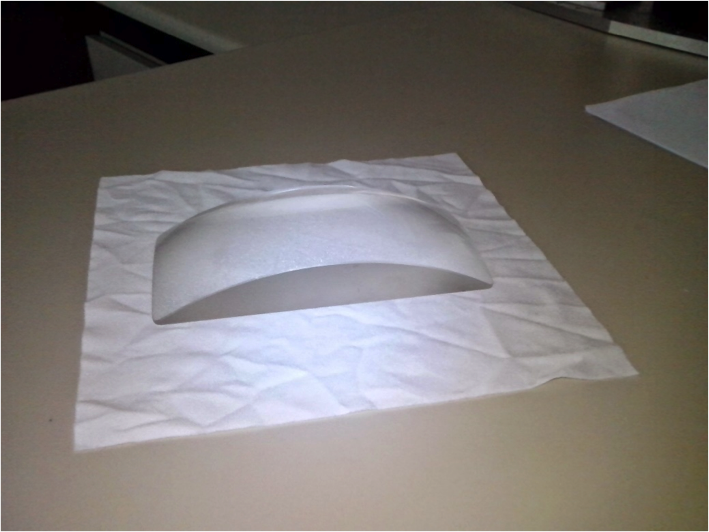}
\caption{The lens used in the testbeam.}
\label{fig:lens}
\end{figure}

\begin{figure}[htbp]
\centering
\includegraphics[angle=90, scale=0.45]{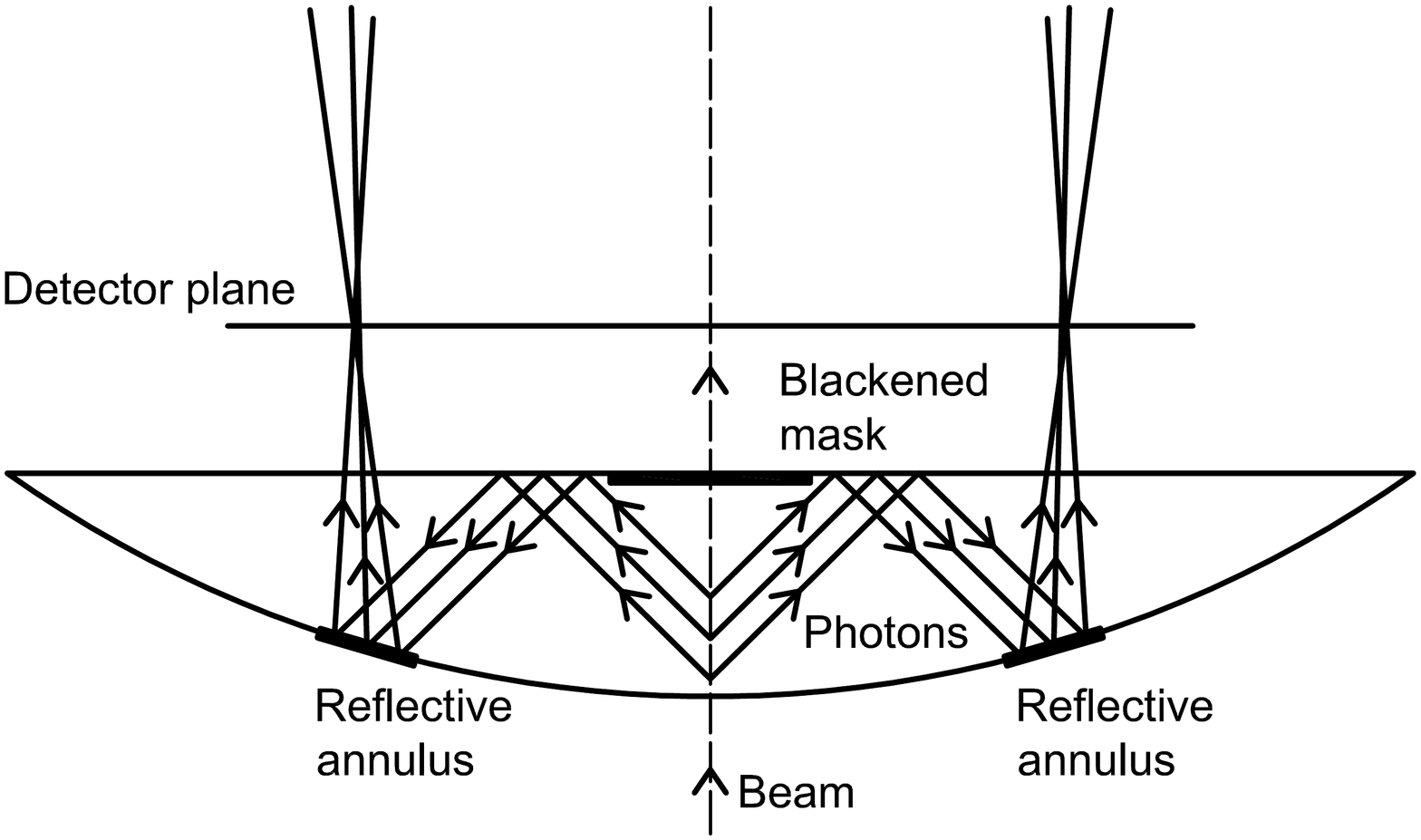}
\caption{Schematic of the optical setup. Particles traverse the
lens generating Cherenkov light, which is reflected at the flat surface and
reflected and focused on the curved surface of the lens. Most rays converge at
the position of the photon detectors}
\label{fig:setup}
\end{figure}

The optical setup consisted of a planoconvex lens made
of borosilicate glass with two parallel cuts giving a doubly-truncated
profile as shown in Figure~\ref{fig:lens}. The radius of
the lens has been measured to be $R=144.6 \pm 0.1 \unitm{mm}$. It has a diameter of
$151.7 \pm 0.1 \unitm{mm}$ and a thickness at the centre of $27.0\pm
0.1 \unitm{mm}$. A reflective layer (annulus) $20\unitm{mm}$ wide was deposited on the spherical surface. The spherical, top and bottom surfaces of the 
  lens have been blackened in order to absorb scattered photons. The centre of the lens on the flat surface, out to a radius of $17\unitm{mm}$ has been blackened to reduce the length of the
path where detectable photons are produced. 
Particles enter the lens at the middle of the
spherical side producing Cherenkov  photons which are reflected at the flat
surface due to total internal reflection. The photons are reflected again by the thin
reflective layer deposited onto the spherical surface and exit from the lens at
the flat surface. At the focal plane of the lens, the photons form a Cherenkov ring as shown in Figure~\ref{fig:setup}. 

\begin{figure}[htbp]
\centering
\includegraphics[scale=0.45, angle=90]{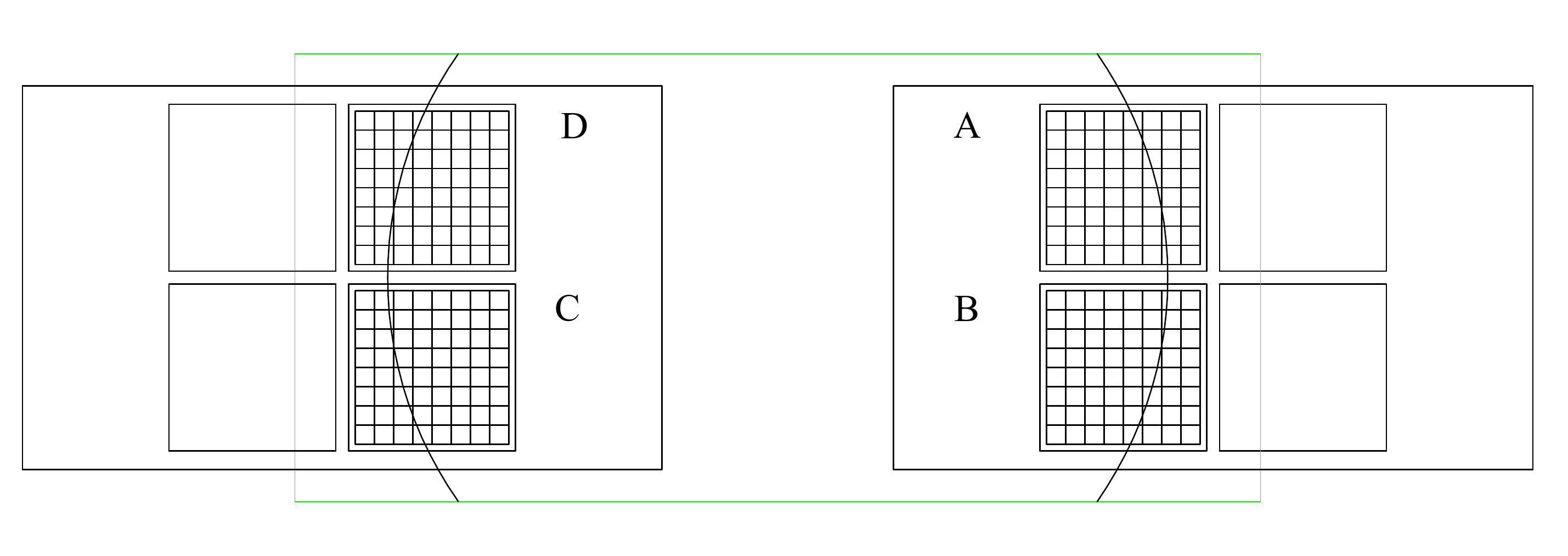}
\caption{The expected ring position superimposed to the two Elementary
  Cells, each with
  two $8 \times 8$ pixel MaPMTs. The beam is entering into the paper.}
\label{fig:ringmapmt}
\end{figure}

A view of the system, looking downstream, with the expected position of the ring on the MaPMTs is shown in Figure~\ref{fig:ringmapmt}. The beam passes through the centre of the image
creating a Cherenkov ring with a radius of $\sim 60 \unitm{mm}$. Only
four of the available 8 MaPMT sockets are instrumented. 

\subsection{Photon detector}
The photon detectors tested are the Hamamatsu R11265~\footnote{http://www.hamamatsu.com/resources/pdf/etd/R11265U\_H11934\_TPMH1336E.pdf},
which are candidates for the RICH upgrade.
The R11265 is an MaPMT capable of detecting single 
photons in the wavelength range 200 -- 600~nm. 
It is a $26.2 \unitm{mm}$ square device, with
$64$ $(8 \times 8$) pixels, and an active area of $23 \times 23 \unitm{mm^{2}}$ 
such that the geometrical acceptance is $77\%$. The pixel size is 
approximately $2.9 \times 2.9 \unitm{mm^{2}}$. The typical average gain 
with a standard voltage divider provided by the manufacturer
is $10^{6}$ at $1\unitm{kV}$. The maximum pixel-to-pixel gain
variation is a factor 3.

A dedicated radiation tolerant ASIC, the CLARO, was used to read out the 
photon detectors. The CLARO is an 8-channel chip containing an analogue 
pulse shaping amplifier and a binary discriminator allowing 
the detection of single photons. Each channel has an individual threshold, programmed through a 6 bits register, resulting in 64 threshold values. 
The baseline of the amplifier is recovered within 25 ns, in order to
suppress signal spill-over, and it has low power consumption at $\sim 1
\unitm{mW}$ per channel. 

\begin{figure}[htbp]
\centering
\includegraphics{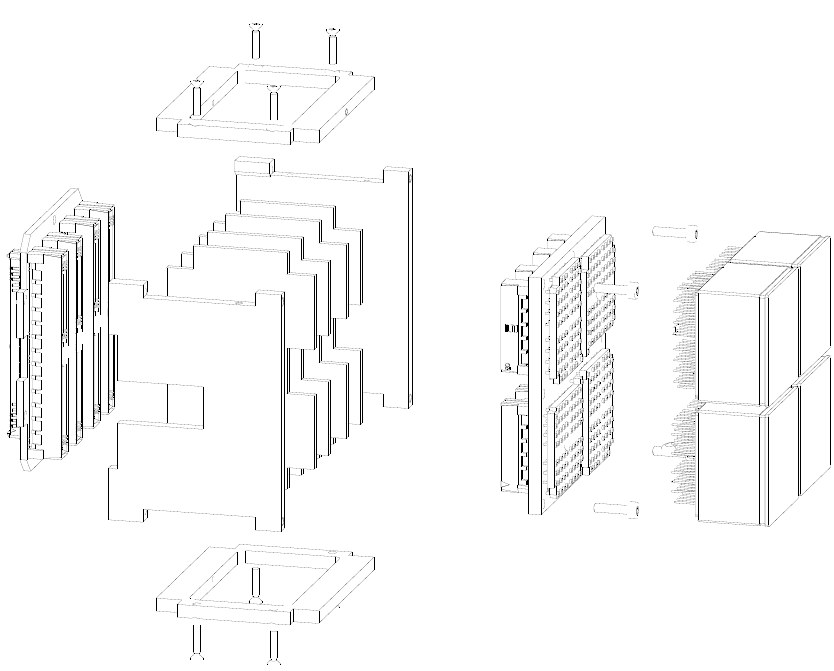}
\caption{An exploded view of the Elementary Cell. Showing from left to right, the BkB, 4 FEBs with aluminium shell, BB and MaPMTs.}
\label{fig:EC}
\end{figure}

The photon detector assembly, housing MaPMTs, readout electronics and
ancillary systems, is conceived as a modular structure based on
independent functional units. The basic unit is the Elementary
Cell (EC) designed to be common to both RICH detectors of LHCb. 
Each EC houses 4 MaPMTs and front-end electronics and consists of:
\begin{itemize}
\item The baseboard (BB) with custom sockets to house 4 MaPMTs. It provides
power, common High Voltage (HV) to the photocathodes of the MaPMTs, 
four resistor divider chains which supply potentials to the dynodes
and connect the MaPMT anodes to the CLARO inputs.
\item Four Front-End Boards (FEB), each equipped with eight CLARO chips.
\item The backboard (BkB), which interfaces the FEBs to the
Digital Board(DB) for configuration and read out.
\end{itemize}
All the components are assembled within an aluminium case
serving as mechanical support structure and providing heat dissipation
and ground connection. 
An exploded view of the EC is shown in Figure~\ref{fig:EC}. 
Two EC have been installed at the focal plane of the lens 
on either side of the beam.
Only half of each EC was instrumented and read out by two FEBs. 
The setup was installed on a remotely controllable translation table 
in order to align it in the plane perpendicular 
to the beam.

\begin{figure}[htbp]
\centering
\includegraphics[scale=0.12]{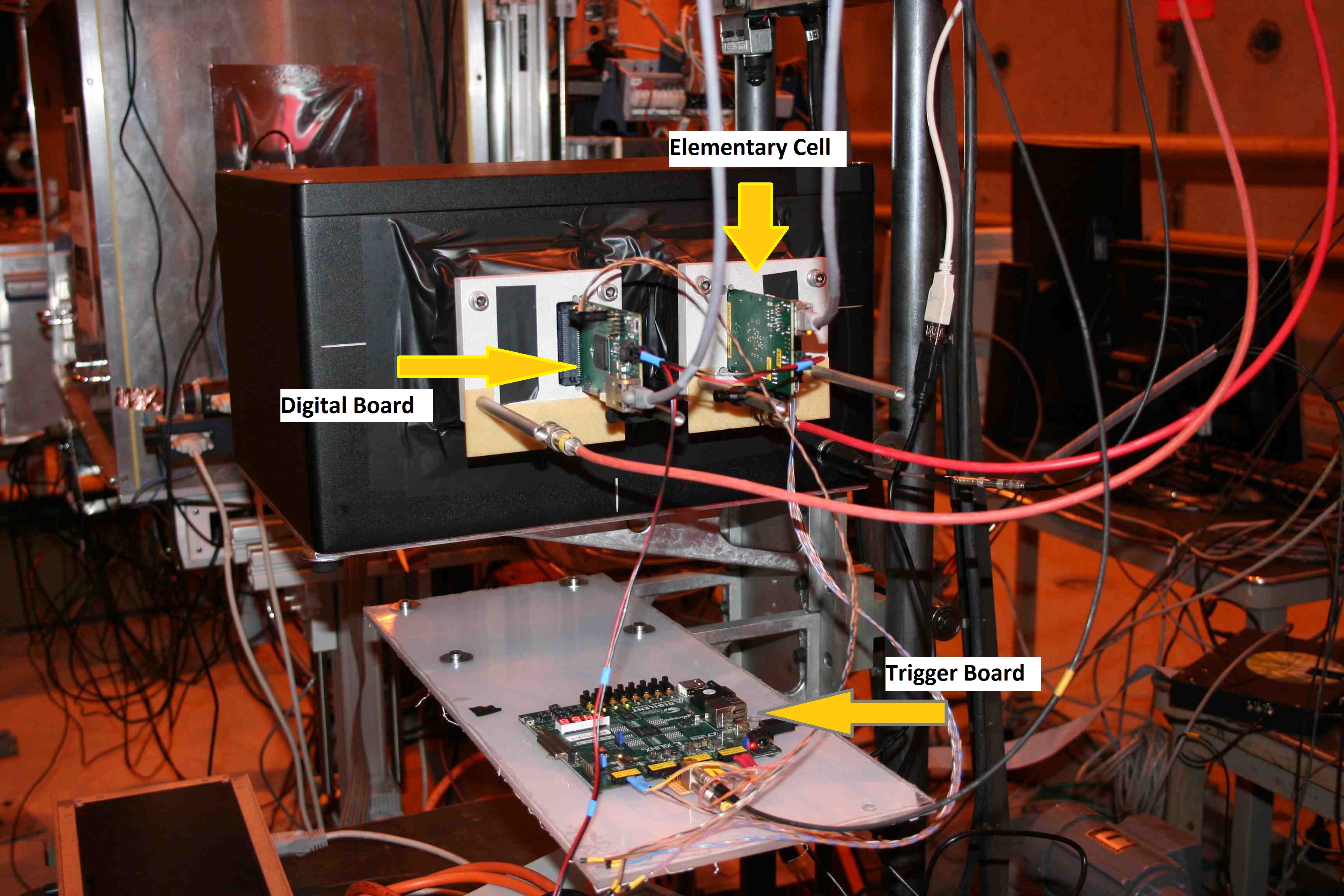}
\caption{The box installed in the testbeam experimental area with two half ECs, equipped with 2 FEBs able to read 128 channels (2
  MaPMTs each).}
\label{fig:box}
\end{figure}

\begin{figure}[h!!!]
\centering
\includegraphics[scale=0.1]{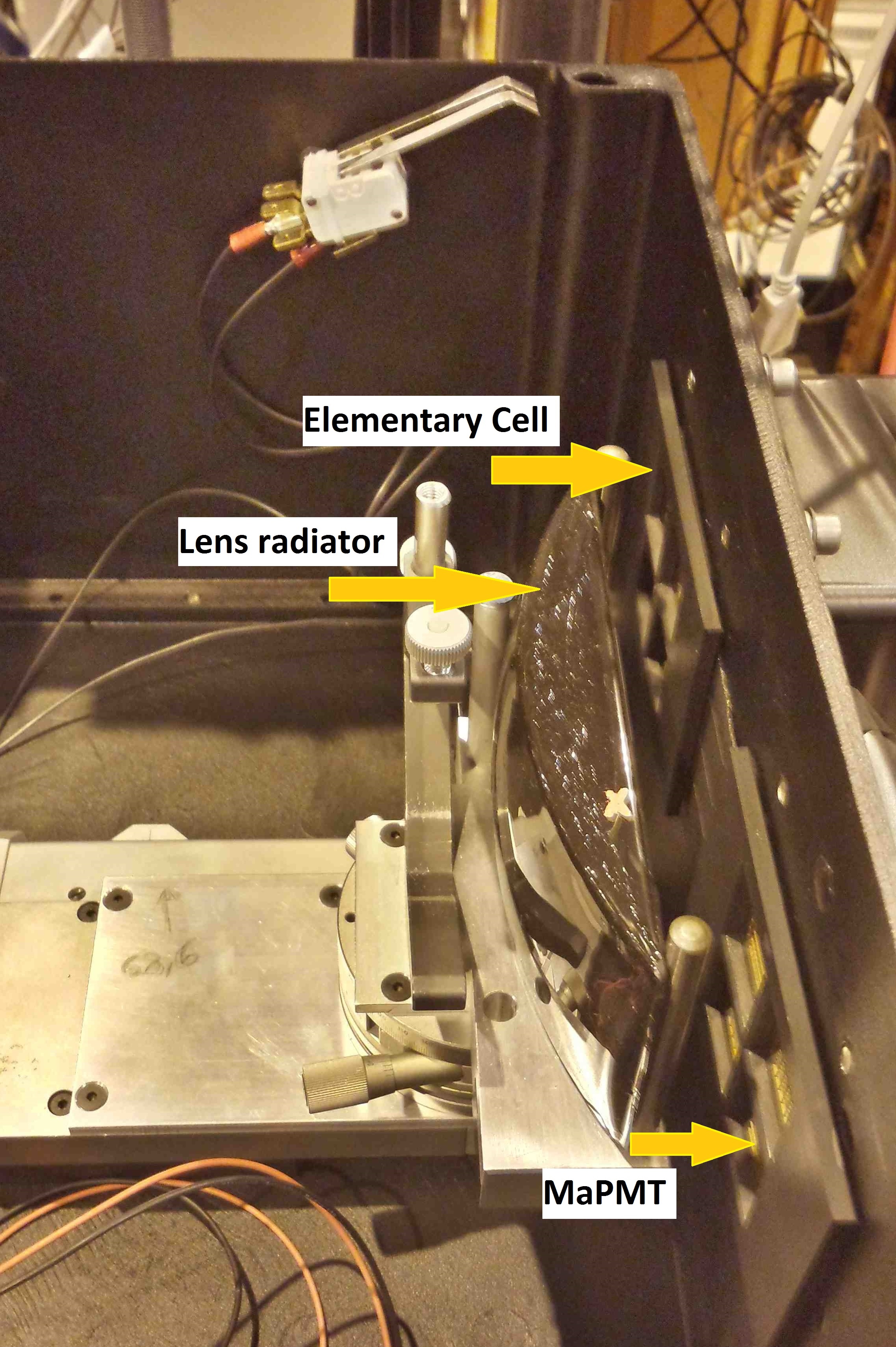}
\caption{A view of the inside of the box. The borosilicate-glass lens
  and the entrance windows of the two ECs are visible.}
\label{fig:inside}
\end{figure}

Figure~\ref{fig:box} shows the light-tight box installed in the experimental area, where 
the two EC are visible.
A view of the inside of the box is shown in Figure~\ref{fig:inside}.

\subsection{Trigger and Data Acquisition}
\label{sec:tdaq}
The digital signal outputs of the CLARO are connected to an FPGA on 
the Digital Board,
collecting events in response to triggers.
The DB formats the data into multi-event packets (MEP) and transmits 
them via Gbit Ethernet to a PC where they are saved.
The Gbit Ethernet links are also used to configure the FPGA on the 
DB and the CLARO chips.
The control of the data aquisition (DAQ) system is achieved using a Graphical User Interface
(GUI). The GUI manages configuration parameters for the EC (such as the CLARO chip
thresholds) and provides run control functions.
Triggers can be generated externally by beam particles or by an internal pulser. 
Particles traversing two scintillator planes placed up and downstream
the beam telescope generate a coincidence that can be used as an
external trigger. This trigger signal is received by the trigger
board, where it is conditioned and fanned out to the
DBs. Alternatively the trigger board can generate its own triggers. The logic on the trigger board also receives a gate signal from
each connected DB indicating that it is ready to accept a
trigger. Triggers are only sent when the gate signals of all connected
DBs are asserted in order to prevent buffer overflow at high trigger rate.
One of the trigger outputs is sent to the beam telescope trigger board
to
provide synchronisation between the two systems.

\subsubsection{Online Data Monitoring}
\begin{figure}[htbp]
\centering
\includegraphics[scale=0.5]{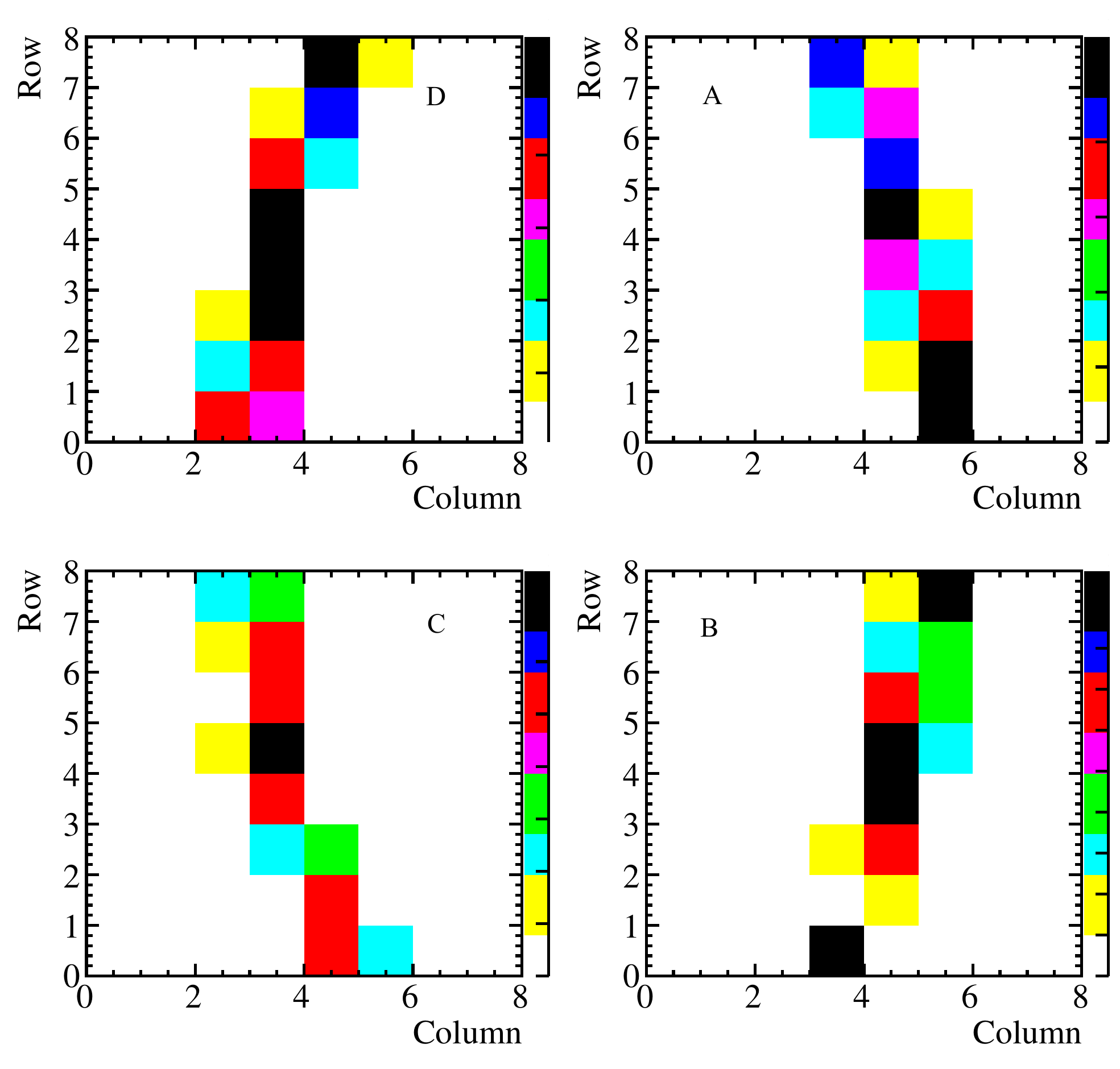}
\caption{A typical map of accumulated hits in a run on the four read out
  MaPMTs. Each MaPMT clearly shows a section of the Cherenkov ring. The
  horizontal distance between the two pairs of MaPMT is not to scale.}
\label{fig:monitoring}
\end{figure}

A quasi-online data monitor has been developed in order to validate 
the data during acquisition. A decoding program is able to read the data 
before the end of the run and to display monitoring histograms.
A map of the accumulated hits on the four MaPMTs is shown in Figure~\ref{fig:monitoring}. Other
histograms, such as distributions of the hit multiplicity, are also
available to monitor the performance of the system.

\subsection{Track Telescope and Readout Synchronisation}
\subsubsection{Telescope Description}
\label{sub:telescope}
The beam can be tracked in space with a dedicated track telescope, 
comprising 8 planes of silicon pixel detectors read out with the TimePix3 (TP3) chip.
Each plane has a size of about $14~\mathrm{mm} \times 14~\mathrm{mm}$ and is
subdivided in pixels of $55~\mu\mathrm{m} \times 55~\mu\mathrm{m}$.
At a typical particle momentum of 180\gevc,
the reconstructed tracks provide excellent position resolution of order 
a few $\mu$m at the lens. The telescope has a triggerless readout: hits are 
recorded continuously once a run is started. Each hit pixel records 
the position of a particle and a timestamp with a precision of 1~ns. 
The data are then analysed offline and tracks are formed from hits in 
each plane that have consistent times.
The telescope also writes 64-bit {\it trigger timestamps}, which are saved
every time an external trigger is sent to the TP3.

The tracking information for the particles passing through the lens radiator
is essential in order to reconstruct the Cherenkov angle and to improve the
simulation by providing it with the correct beam profile.
Moreover, the tracking data can identify events with multiple beam particles
within the trigger gate of the MaPMT acquisition and help correctly estimate
the number of expected photons per event.

\subsubsection{Synchronization between the two systems}
\label{sub:synctelrich}
\begin{figure}
\centering
\includegraphics[width=0.48\textwidth]{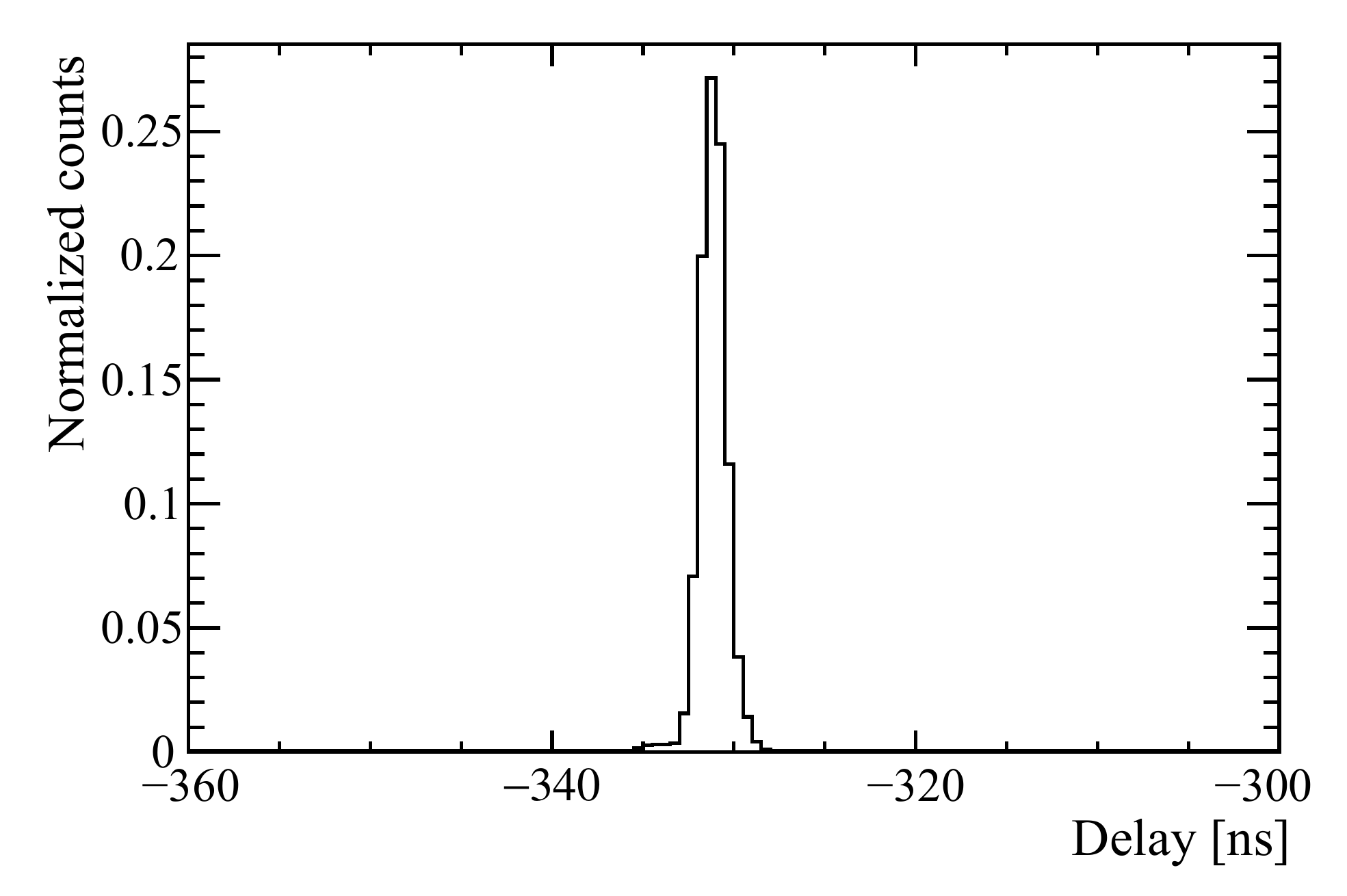}
\includegraphics[width=0.48\textwidth]{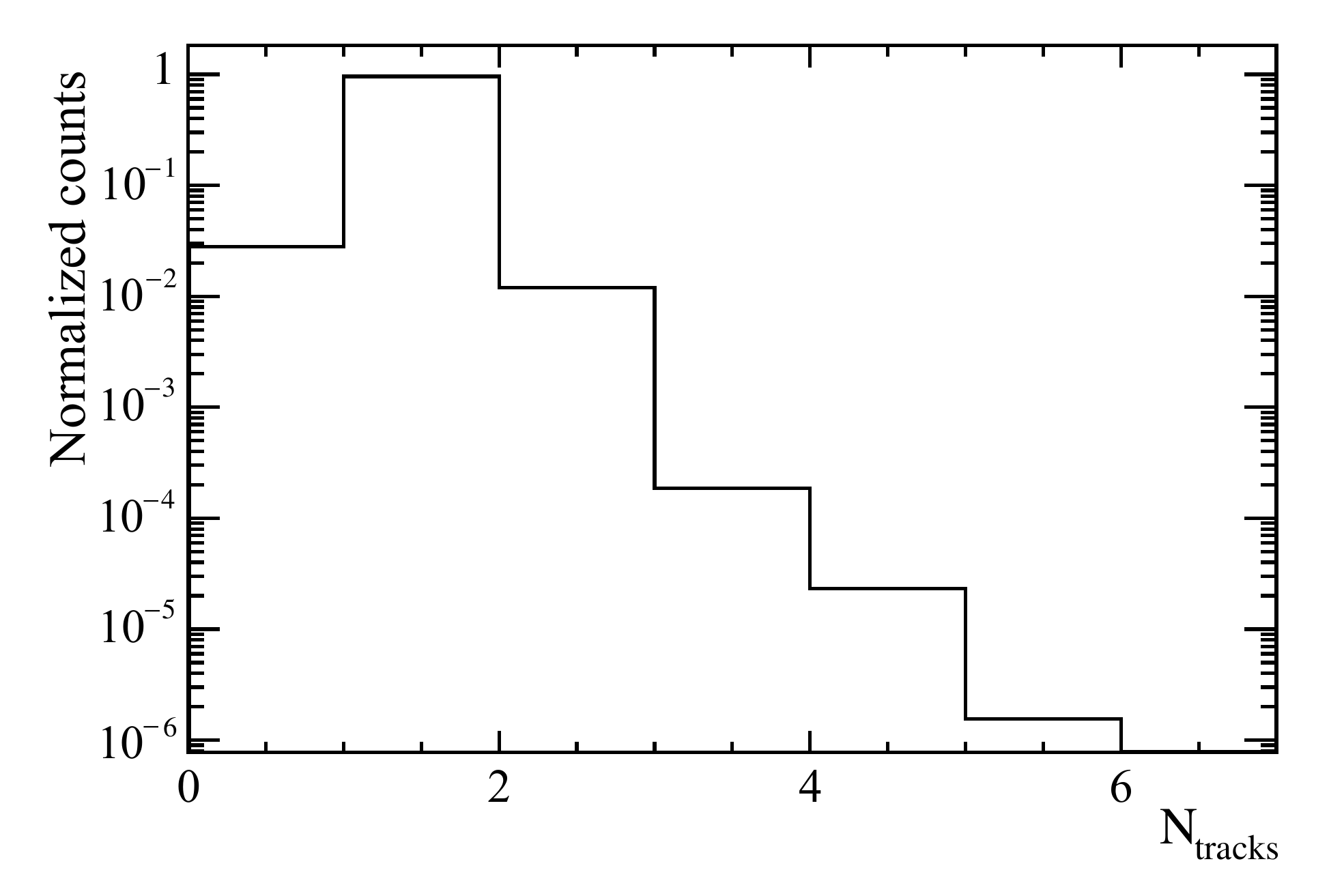}
\caption{Left: Distribution of the delay between RICH triggers and
  tracks recorded by the telescope. Right: Distribution of the number of tracks associated to about $6 \times 10^5$ events. Less than $3\%$ of the events have no tracks associated, while about $1\%$ of the events have more than one track associated.}
\label{fig:tel}
\end{figure}

The association between the hits collected with the RICH acquisition system and the tracks
from the telescope is made using timestamps.   
Every time a RICH event is acquired a signal is sent to the telescope, where it is registered and 
time stamped. There is a time delay of about 330~ns  between the RICH signal and the tracks, 
as shown in Figure~\ref{fig:tel}. The search time window was a few hundred ns.
The time distribution of the tracks is very sharp,
with an RMS of 2~ns showing good communication between the two systems.
Figure~\ref{fig:tel} shows the number of tracks that can be found
within a 50 ns window centred at -330 ns. More than 97\% of the RICH events have at least one
telescope track associated with the event, while around 1\% of the events have more
than one associated track. The missing track information, in less than 3\% events, is due to the inefficiency of the tracking system. 

\section{Simulation}
\label{sec:simulation}
A simulation was used in order to understand and
optimise the experimental setup. Two different methods have been used
to allow quick development and to cross check the results.
A ray-tracing simulation using a fast and flexible optical CAD software 
package\footnote{Optica 3, http://www.opticasoftware.com and Wolfram Mathematica 9,
http://www.wolfram.com/mathematica} has been used to optimise the optical
system and make an initial estimate of its performance.  In addition to this, a
more detailed simulation of the testbeam setup using the 
\geant~\cite{Agostinelli:2002hh} software toolkit was
performed for a complete evaluation of the system performance, including the
reconstructed Cherenkov angle resolution.

\subsection{Optical simulation}
In preparation for the beam test the optical CAD software 
has been used to assess different setup configurations. The goal was to find a
configuration able to focus the Cherenkov photons on a focal plane which is
sufficiently displaced from the beam axis to avoid direct exposure of the MaPMTs
to the particles of the beam. 
The measured transmission properties of the used lens were inputs for the simulation.

\begin{figure}[htbp]
\centering
\includegraphics[scale=0.6, angle=-90]{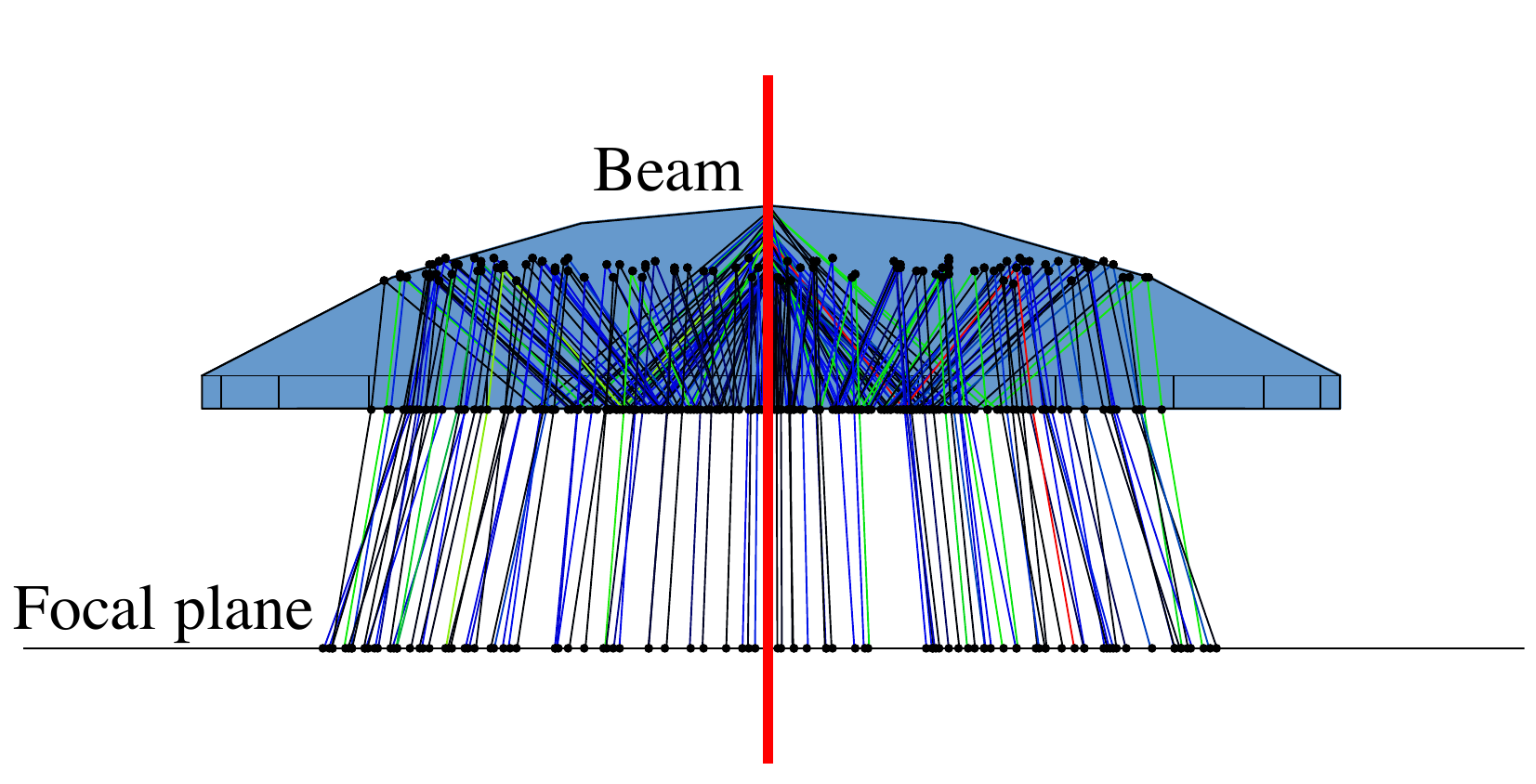}
\caption{The Cherenkov photon tracing (seen from the top) obtained using the optical
  CAD. The different colours represent Cherenkov photons with different
energy.}
\label{fig:raytracing}
\end{figure}

\begin{figure}[h!!!]
\centering
\includegraphics[width=7cm]{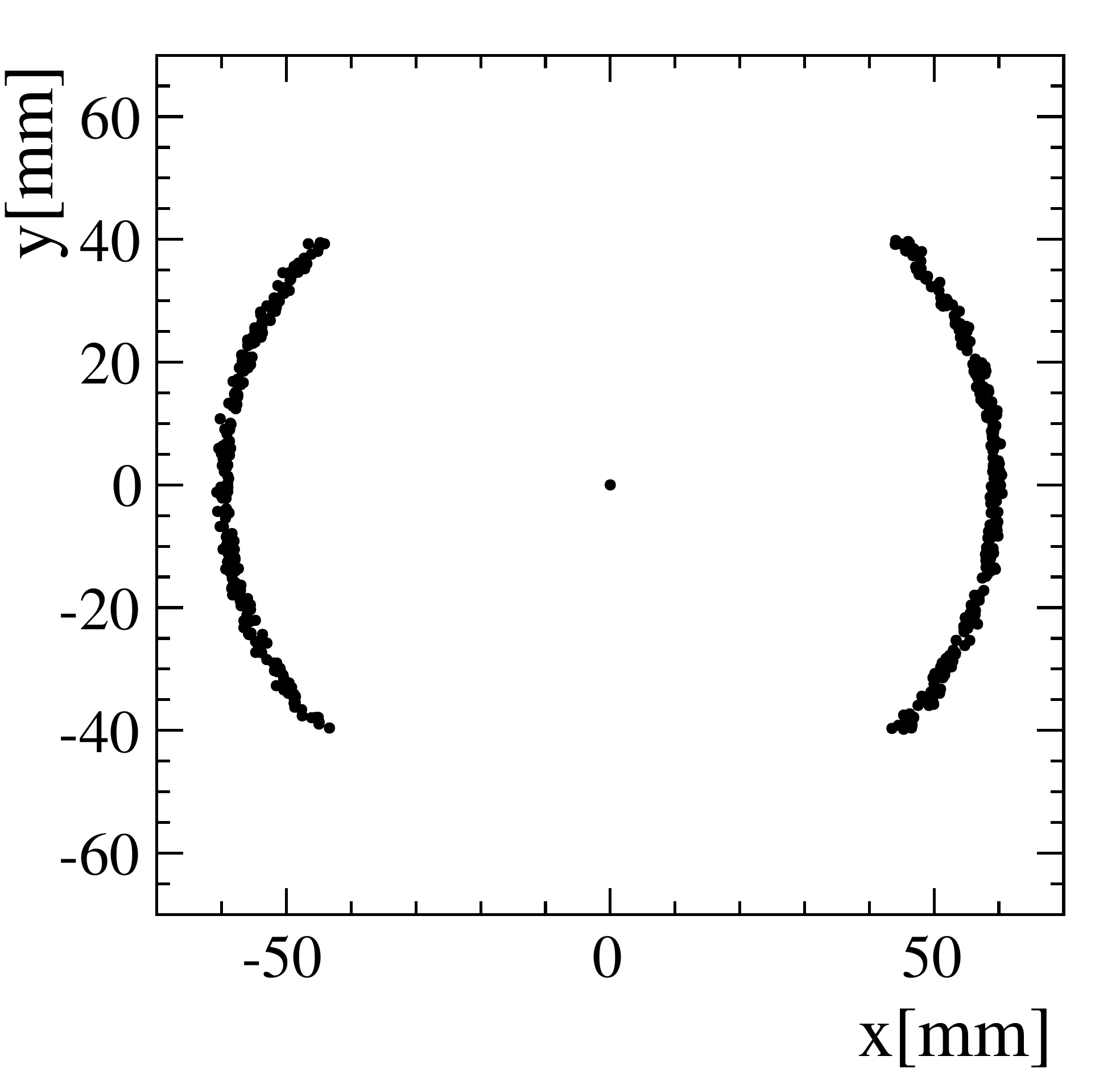}
\includegraphics[width=7cm]{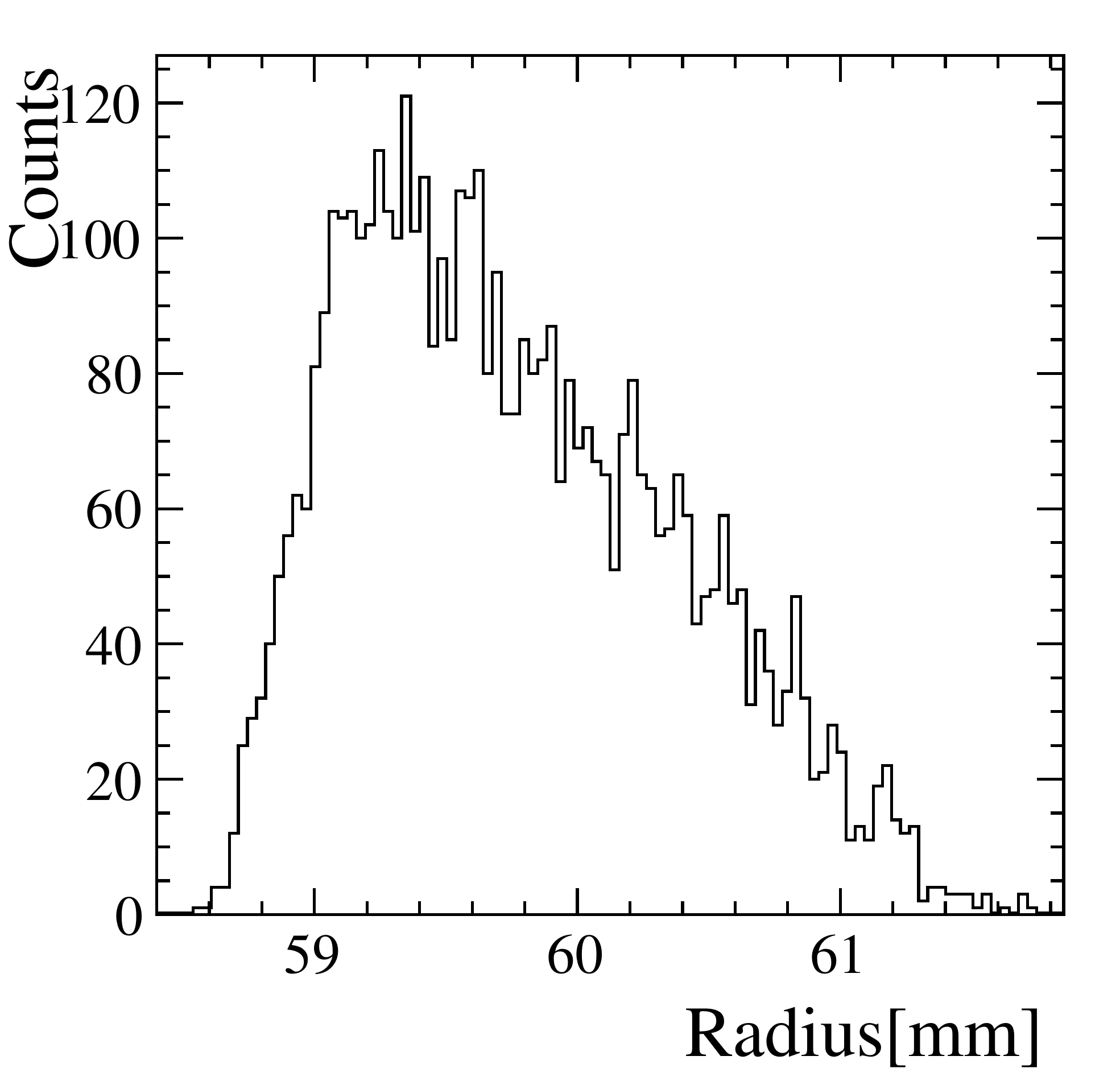}
\caption{Expected ring of Cherenkov photons at the photodetector
  plane on the left. On the right, the radius distribution with an RMS of $0.6\unitm{mm}$. Since the lens is cut at the top
  and bottom, the expected ring is cut at $\pm 40 \unitm{mm}$ in
  the vertical direction.}
\label{fig:ring}
\end{figure}

Figure~\ref{fig:raytracing} shows ray-traced photons that correspond to the
geometrical setup shown in Figure~\ref{fig:setup}. 
Figure~\ref{fig:ring} shows the Cherenkov ring, which is incomplete as the
lens is cropped and does not cover a full circle. The thickness of the ring
is of the order of one MaPMT pixel, less than 3\unitm{mm}. 
The best focal distance is determined by simulations and located at $25\unitm{mm}$ from the flat side of the lens. The focused ring has an
expected radius with a mean value of $R_{\rm ring} = 59.8\unitm{mm}$ and an RMS of $0.6\unitm{mm}$. (Figure~\ref{fig:ring}, right).

\begin{figure}[h!!!]
\centering
\includegraphics[width=0.45\linewidth]{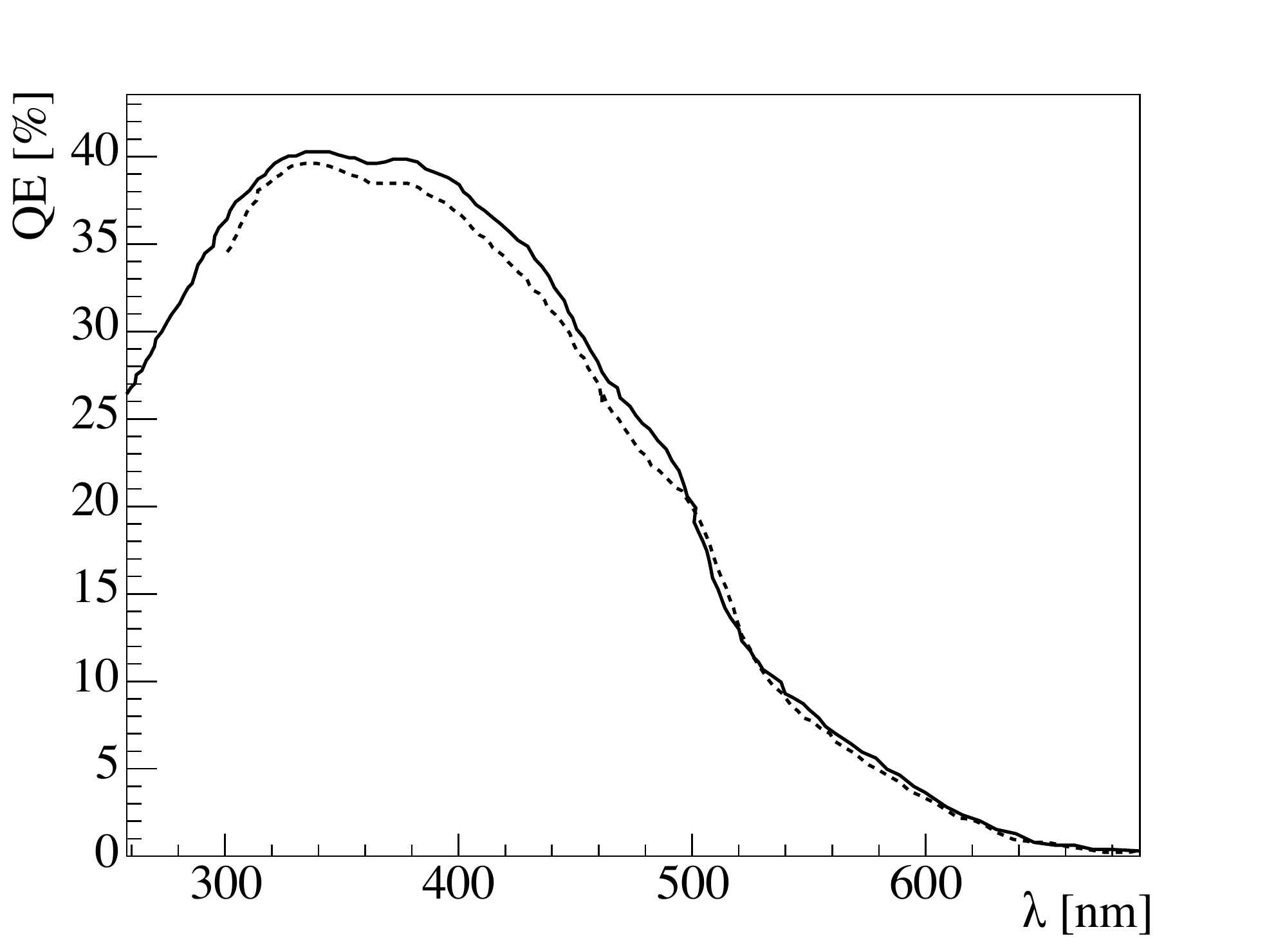}
\includegraphics[width=0.45\linewidth]{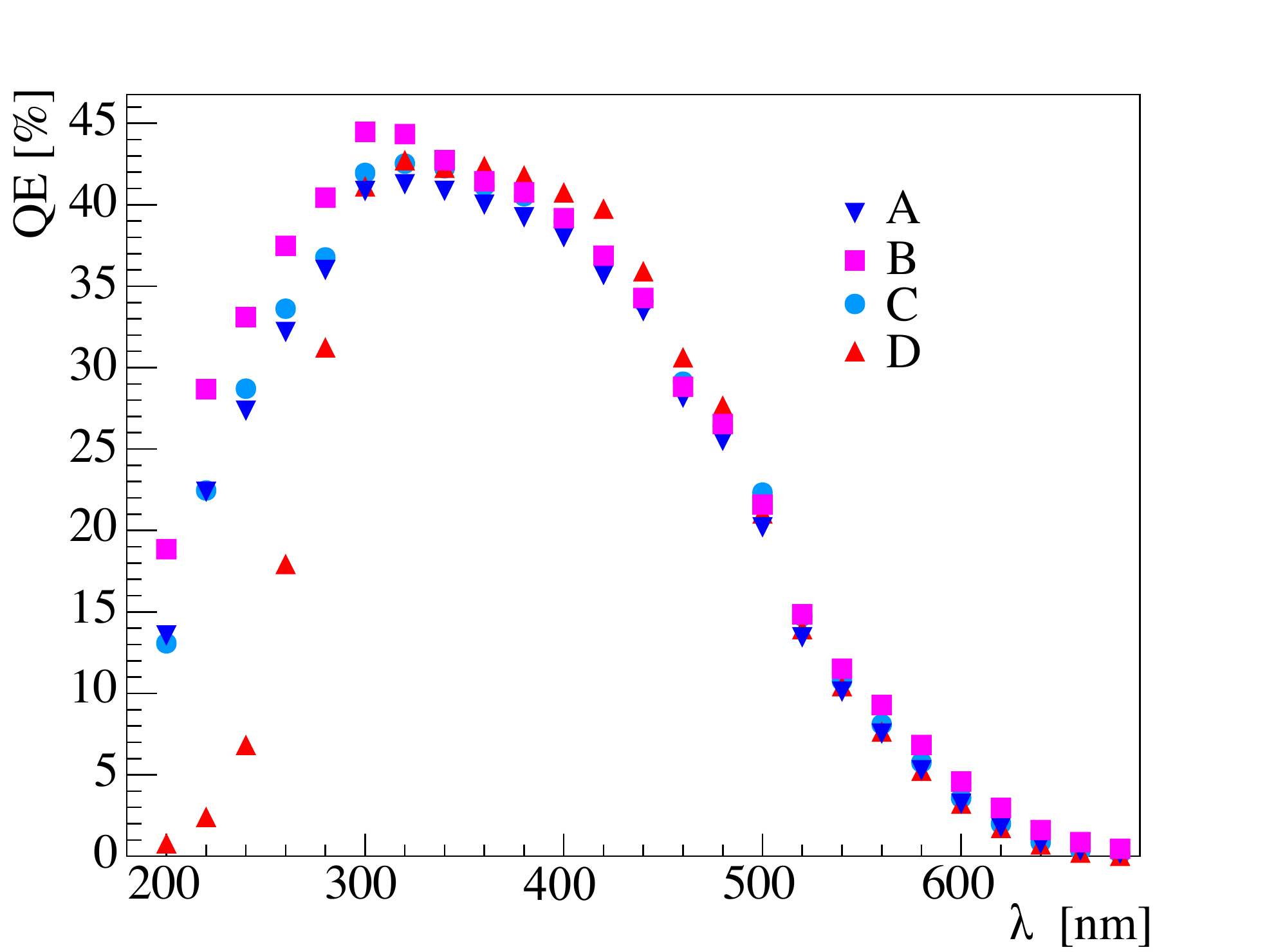}
\caption{The typical Super Bi-alkali photocathode quantum efficiencies
  from the tube manufacturer for borosilicate (BS) glass (dashed) and UV
  glass (continuous) on the left. On the right, the measured quantum efficiencies for
  the four MaPMTs used in these tests.}
\label{fig:sr}
\end{figure}

The thickness of the ring was minimised by studying the effects of the 
emission point of Cherenkov photons in the radiator and the chromatic dispersion. In order to take into
account the chromatic dispersion, the full Cherenkov spectrum was generated and then convoluted with
the quantum efficiency of the MaPMTs shown in Figure~\ref{fig:sr}. Including both effects
in the simulation and collecting photons from the full length of the
particle path inside the glass, the RMS of the photon distribution
(ring thickness) was 1.9\unitm{mm}. In order to reduce it, black tape was
used to allow only photons emitted in the first 13\unitm{mm} of the glass
to reach the photon detectors. In this way the RMS was reduced 
to 0.63\unitm{mm}. Reducing the number of photons also lowered the
probability of double hits on the same pixel that could bias the
photon yield estimate. 

The contribution of the chromatic dispersion was evaluated by fixing the emission point, giving a contribution to the
RMS of 0.6\unitm{mm}. Similarly, fixing the wavelength of the
emitted Cherenkov photons, the emission point error contribution is evaluated to be $0.12\unitm{mm}$.

Simulating a beam with a Gaussian spread of about 10\unitm{mm} in both
directions had negligible difference compared with a beam with zero
spatial extent. Finally, the biggest contribution to the width of the
ring is the pixel size of the MaPMTs
which at 3\unitm{mm} gives an RMS of 
$\sim\frac{3\unitm{mm}}{\sqrt{12}}=(\sim 0.9\unitm{mm})$.

\subsection{\geant simulation}
The simulation and reconstruction of the testbeam data use a common software
framework so that the same detector description is used for both. It is
configured within the LHCb software framework and uses the \geant
toolkit for the detector simulation. The
information relating to the geometry, material and optical properties
of the various components of the testbeam setup is read from a
database which includes
the size and shape of the radiator, the geometry of the MaPMTs and the pixel size.
The \geant toolkit simulates the physics processes of the charged particle interactions with the
detector, the production and transport of Cherenkov photons, the reflection and
refraction at the optical boundaries and the production of photo-electrons.  
The measured quantum efficiencies of 
the MaPMTs as a function of the photon wavelength are also included
(Figure~\ref{fig:sr}) together with the optical properties
of the glass radiator.  The refractive index is then scaled using the
mean Cherenkov angle reconstructed from the data, as described in the
following section. The beam divergence is obtained from the measurement of the
beam direction using the tracking telescope described in Section~\ref{sub:telescope}. The detection
efficiency of each of the pixels in the MaPMTs, due to the readout threshold
applied, is obtained from the measurements taken during the testbeam.    The
number of detected photons  per charged track obtained from the simulation is
compared with that from data in Section~\ref{sec:dataSamples}.

%

%
\section{Data analysis}
\subsection{Threshold scan studies}
\label{subsec:threshold}
\begin{figure}[b!!!]
\centering
\includegraphics[width=\linewidth]{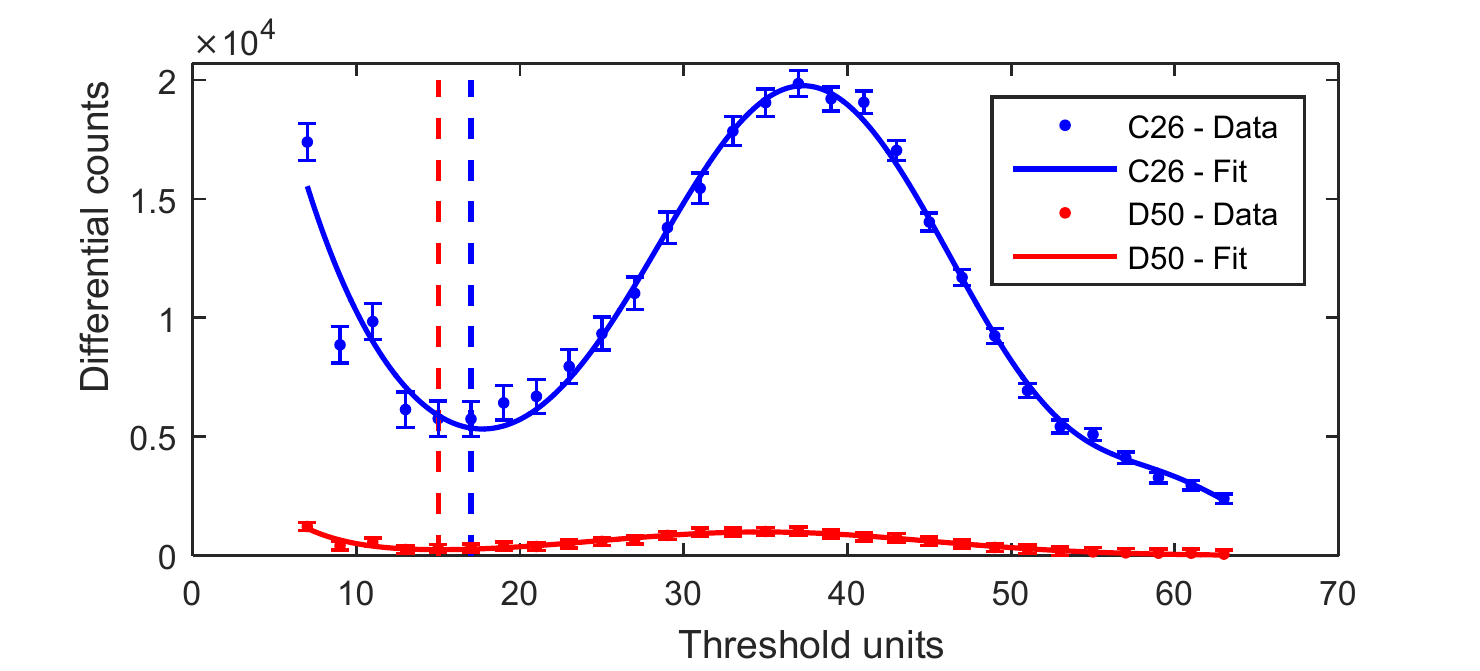}
\caption{Single photon spectrum for a pixel on the ring (pixel 26 MaPMT C) in blue (upper curve) and for a pixel off the ring (Pixel 50 MaPMT D) in red (lower curve). The dashed lines indicate the chosen value of the threshold for each pixel (17 for pixel C26 and 15 for pixel D50).}
\label{thrscn:spectrum}
\end{figure}

\begin{figure}
\centering
\includegraphics[width=.8\linewidth]{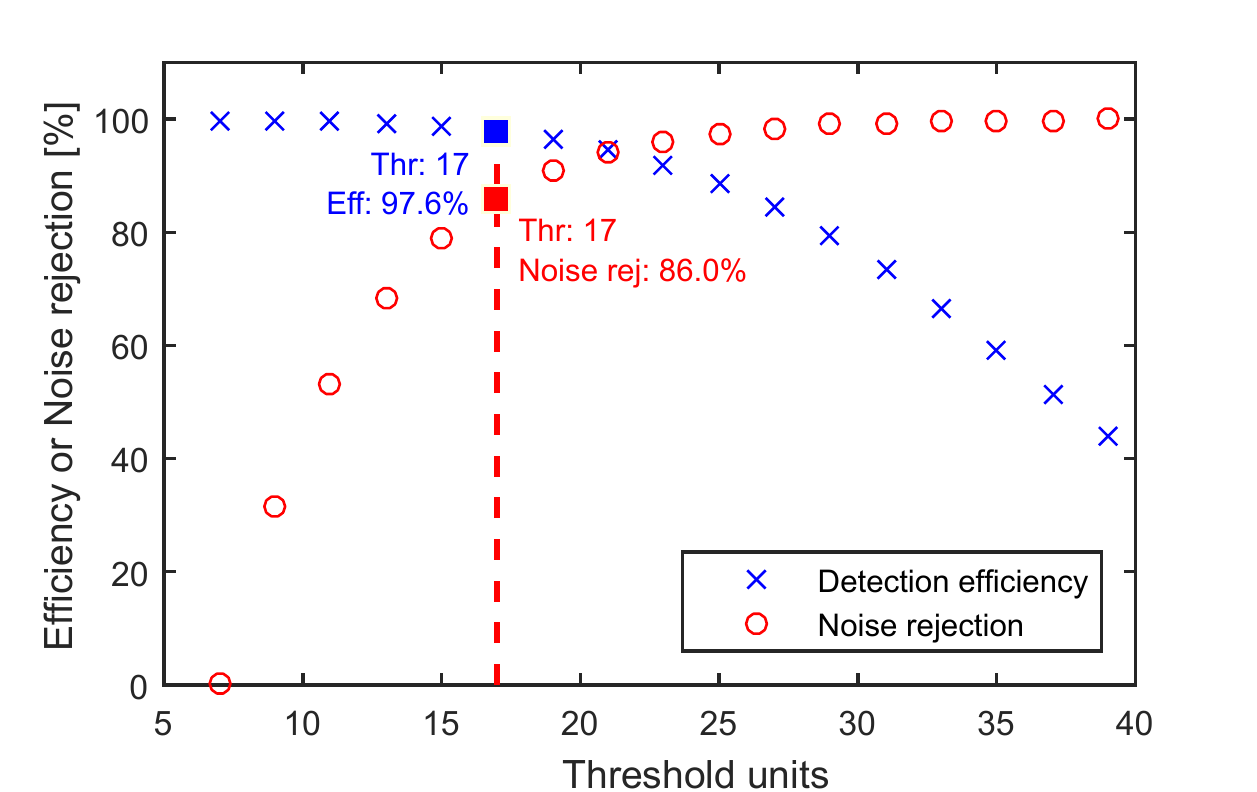}
\caption{Single photon peak detection efficiency and noise rejection vs threshold setting for pixel 26 on MaPMT C. The red line and the data-tips indicate their value at the chosen threshold setting.}
\label{thrscn:efficiency}
\end{figure}

Several threshold scan runs were performed throughout the test period, at various MaPMT bias voltages, all with 1 million events per threshold setting. The threshold settings spanned from threshold 7 to threshold 63 in steps of 2 units, where each unit is about $35\times 10^3\,e^-$. Threshold scans were performed using Cherenkov photons and this implies that off-ring pixels were illuminated only by stray photons and their event rate is much lower than that for pixels on the Cherenkov ring. Nevertheless the spectra of off-ring pixels could be reconstructed adequately despite the lower statistics.
The integral spectra obtained from these runs were then differentiated and the resulting pulse-height spectra for each pixel were fitted with three Gaussian functions (noise pedestal, single photon peak and double photon peak). Two typical spectra, with the fit superimposed, for a pixel on the ring (blue) and off the ring (red) in the default working condition (MaPMT biased at 1000 V) are shown in Figure \ref{thrscn:spectrum}.
The spectra were used to determine the best value of the threshold for each pixel, defined as the minimum between the noise pedestal and the single photon peak. By selecting this threshold value, it is possible to reject most of the noise while maintaining an adequate photon detection efficiency. Figure~\ref{thrscn:efficiency} shows a plot of the photon detection efficiency and the pedestal noise rejection with respect to the threshold settings for one of the pixels of Figure \ref{thrscn:spectrum}. The efficiency is calculated as the ratio of the integral above threshold to the integral of the complete single photon peak, while noise rejection is calculated as the ratio of the integrated noise pedestal below the chosen threshold to the integral of the pedestal above the lowest available threshold.

\subsection{Measurement of dark counts}
\label{sec:darkcounts}
\begin{figure}
  \centering
  \includegraphics[width=0.6\textwidth]{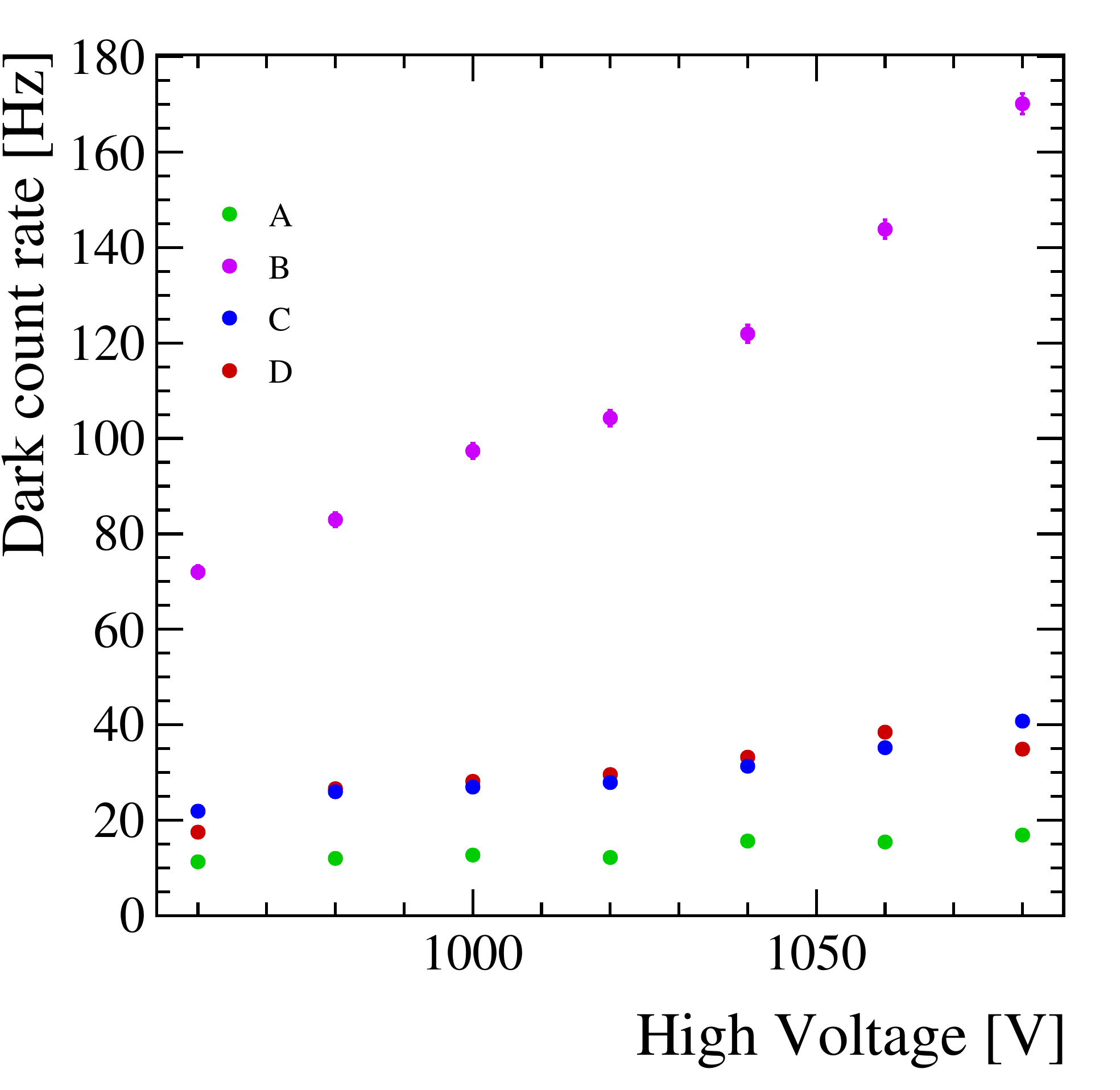}
  \caption{A plot of average dark count rate for each MaPMT as a function of high voltage. }
  \label{fig:DCHVgraph}
\end{figure}

Several runs were taken without beam using the internal trigger described in Section~\ref{sec:tdaq}.
The number of hits recorded by each anode without beam is referred to as the dark count.
Around 10 million events were recorded for each run at different values of high voltage from 960~V to 1080~V in steps of 20~V.
The threshold was set to be uniform across all channels, without using
the individual setting, therefore the high voltage dependence cannot be properly characterised.
In order to estimate the dark count rate, the number of recorded hits at each anode is divided by the number of triggers and the length of time that the device is actively recording photons after each trigger, which was set to $62.5$~ns.
The dark count rate is calculated for all anodes, and the mean value for each MaPMT is plotted as a function of high voltage in Figure \ref{fig:DCHVgraph}.
It should be noted that the dark count rate is negligible compared to the hit rate for runs taken with beam.

The manufacturer provides measurements of the dark current and average gain for each MaPMT, which can be used to calculate the average dark count rate across all anodes in the device.
As the dark current and dark count measurements are made in a very different way, it is only possible to make a qualitative comparison.
The measurements are compatible to within an order of magnitude, i.e. tens of Hz.
The MaPMTs with the highest and lowest dark currents given by the manufacturer, also have the highest and lowest dark count rates as measured in our testbeam setup.

\subsection{Cross-talk studies}
\label{subsec:crosstalk}
From the acquired data it is possible to estimate the cross-talk of the entire opto-electronic chain. This can be done by looking at pairs or clusters of neighbouring active pixels during the same event in a region of the MaPMT far from the Cherenkov ring. On the ring the rate of Cherenkov photons is quite high (about 4 photons per event per MaPMT are expected) so there would be a non negligible probability of two real photons hitting neighbouring pixels, resulting in a false cross-talk count. For this reason all the pixels on the ring and the nearest neighbours are masked out in this analysis. Furthermore, since the illumination rate of pixels off-ring is less than 1\% of the total events, the probability of accidental coincidence between two neighbouring pixels can be neglected. Also the dark counts are low enough to be ignored. For these reasons, the coincidences are attributed to cross-talk. 

Using the binary data from the read out system it is not possible to distinguish which of the two neighbouring pixels induced cross-talk on the other, thus the number of cross-talk events was evenly split between the two pixels.
Based on these considerations the cross-talk probability is calculated as:
\begin{equation}
{\rm \operatorname{Cross-talk}} \left(i\rightarrow j\right) = \frac{N_{ij}/2}{N_i}
\end{equation}
where $N_{ij}$ is the number of events where both pixels are on and $N_{i}$ is the number of events where at least pixel $i$ is on. The same calculation is done exchanging pixel $j$ with pixel $i$ and the final cross-talk value between the pair is the mean of the two values.

\begin{figure}
\centering
\includegraphics[width=.9\linewidth]{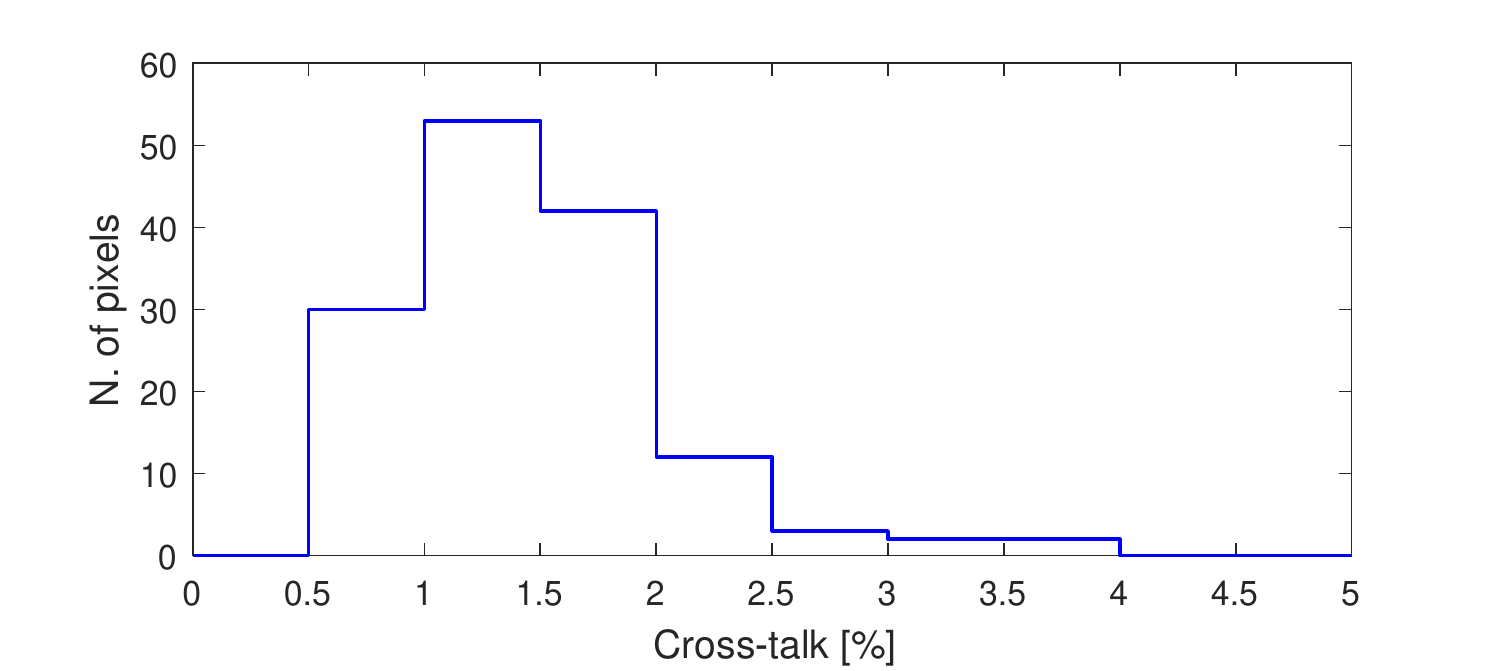}
\caption{Histogram of the cross-talk probability. The mean value is 1.48\%.}
\label{ct:histogram}
\end{figure}

The histogram of the computed values for the runs with thresholds set for each channel, is plotted in Figure~\ref{ct:histogram}. The mean value across the two MaPMT considered (C and D) is 1.48\%. These results are obtained in real working conditions and include the contributions from the whole opto-electronic chain. This low cross-talk was possible thanks to the careful design of the entire system, including the good quality of the MaPMT selected, the layout of the front-end board and the CLARO preamplifier design~\cite{Carniti:2012ue}. This result is in good agreement with previous test-bench measurements.

\subsection{Photo-electron yield measurement}
\label{sec:dataSamples}
\begin{figure}[htbp]
\centering
\includegraphics[scale=0.35]{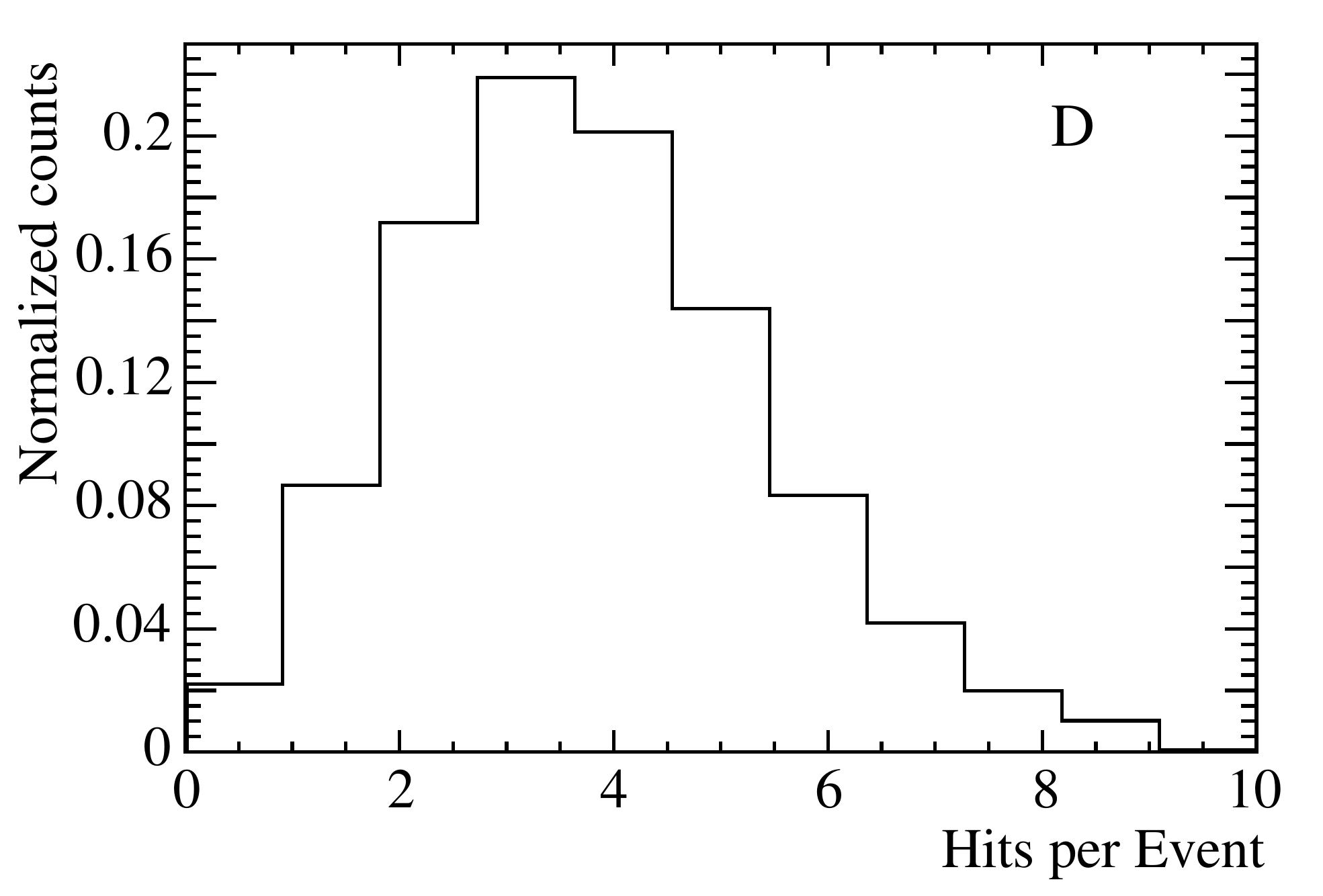}
\includegraphics[scale=0.35]{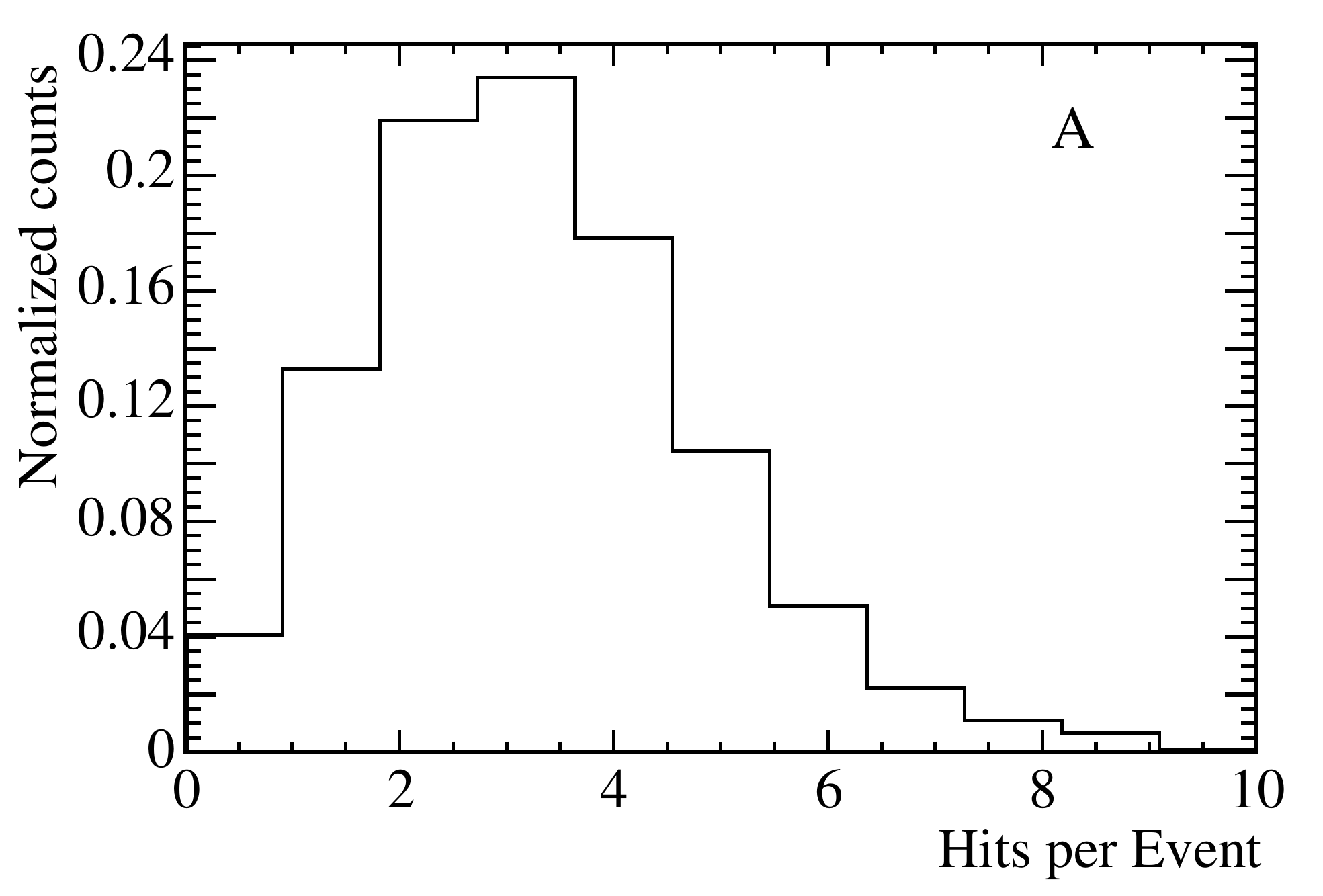}\\
\includegraphics[scale=0.35]{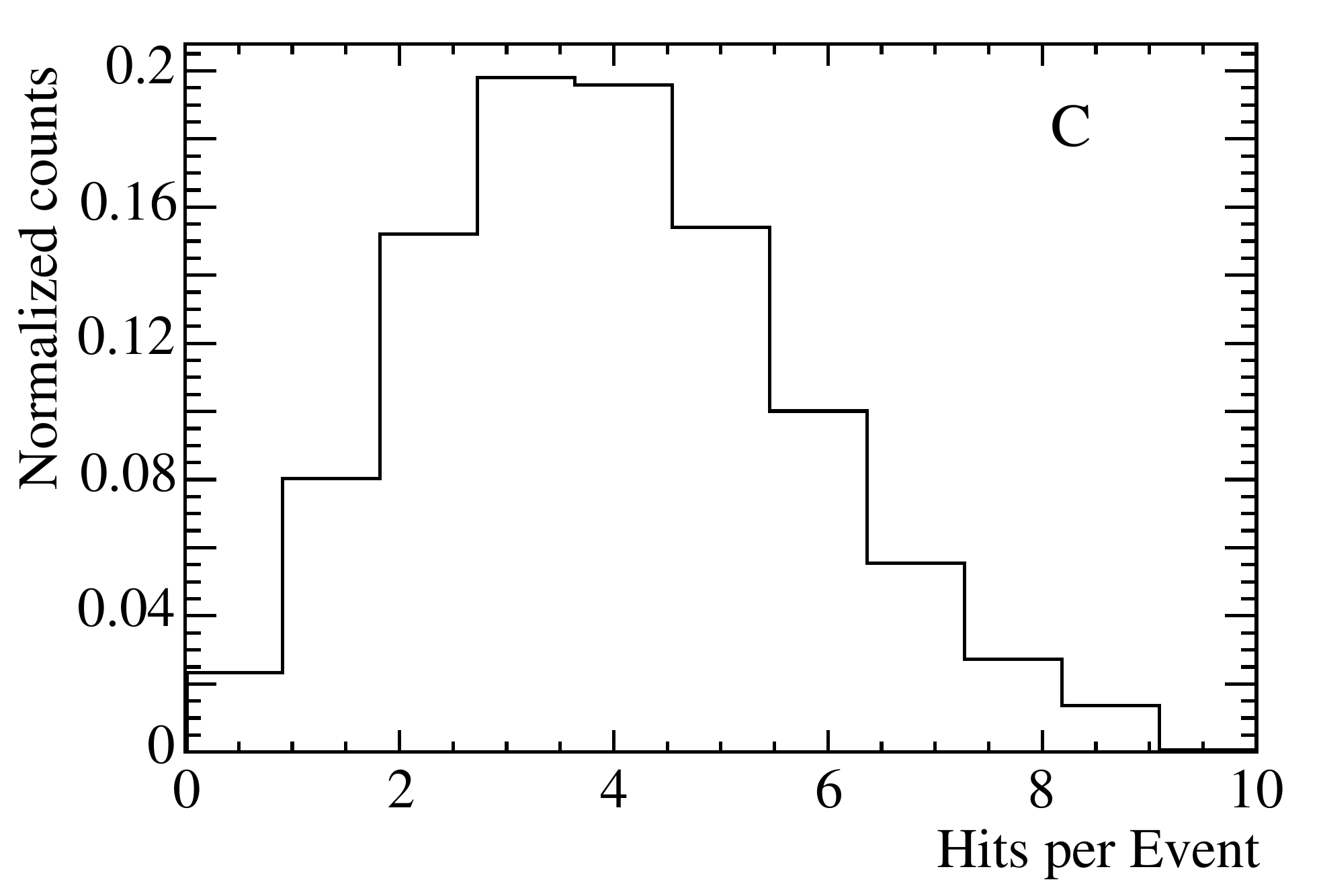}
\includegraphics[scale=0.35]{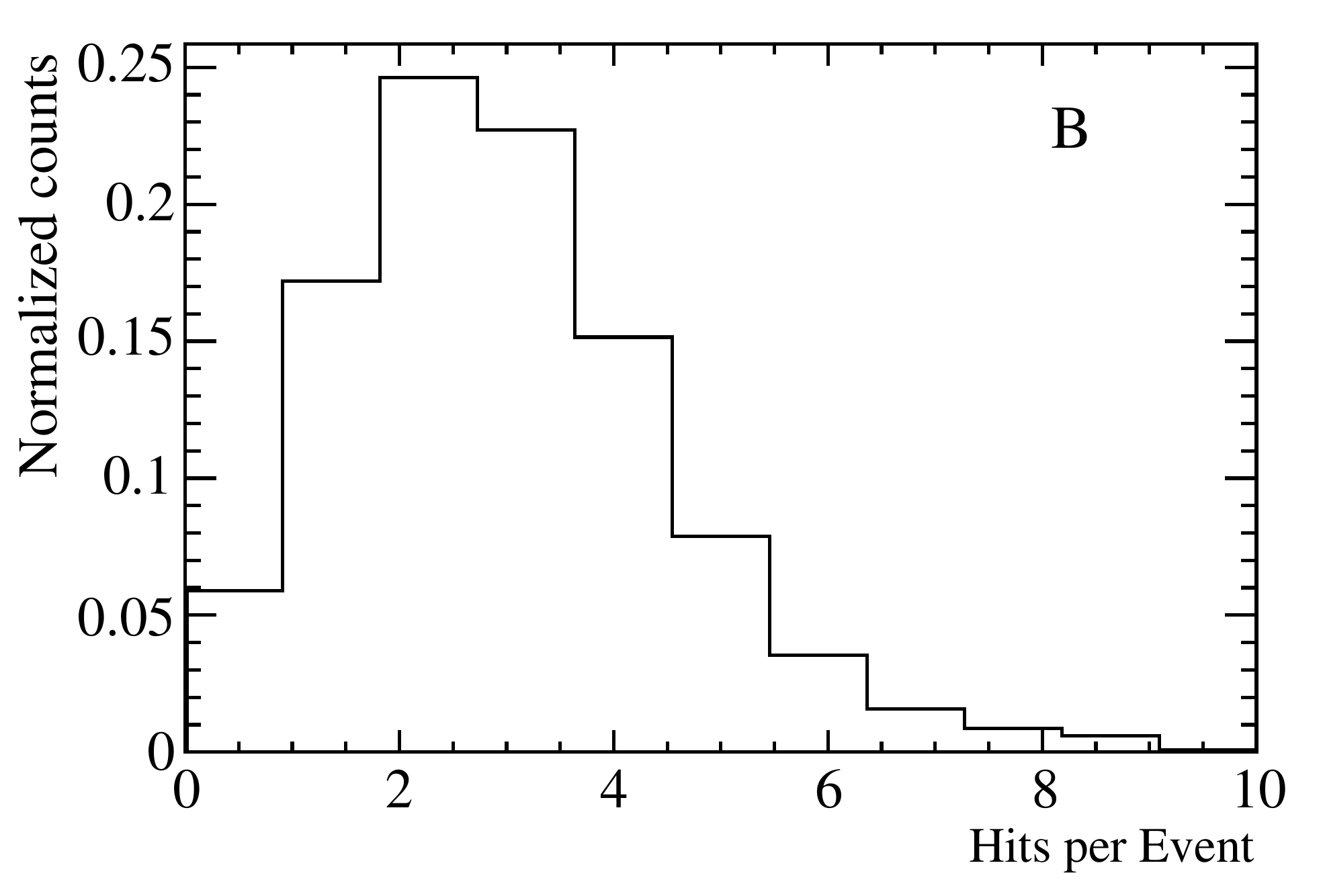}\\
\includegraphics[scale=0.5]{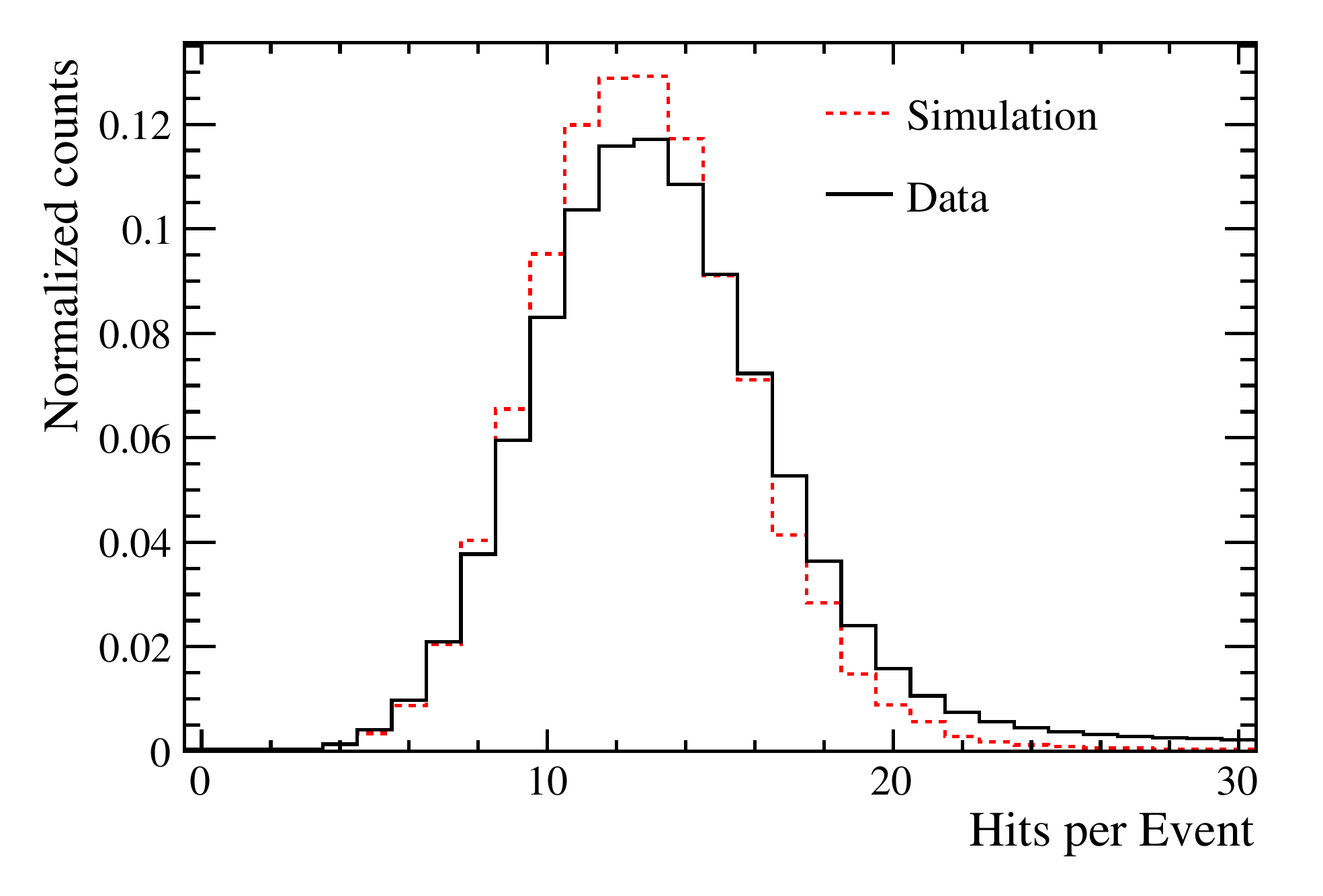}
\caption{Multiplicity distribution for each of the four MaPMTs (top). Distribution of the total number of hits in data with the expected distribution obtained from the \geant simulation superimposed (bottom).}
\label{fig:nhits}
\end{figure}

The distribution of the number of recorded hits has been studied with data
in order to calculate the number of detected Cherenkov photons. A loose
selection has been applied in order to reject possible
noisy events. In particular events with more than 10 hits in any one MaPMT
or more than 30 hits in the four MaPMTs have been rejected. The 
distribution of the number of hit
pixels per event for each MaPMT and in total is shown in Figure~\ref{fig:nhits}.
These are compared with the results from the \geant simulation and are also cross-checked using an analytical estimate.

\subsubsection {Comparison with simulation }
The distributions of the number of hits per track in each MaPMT are shown in the upper plots of Figure~\ref{fig:nhits}. The lower plot compares the total number observed in data with that from the simulation as described in Section~\ref{sec:simulation}. The mean number of hits per track, in data and in simulation, are given in Table~\ref{tab:PhotonYield} and agree within uncertainties.
The uncertainties mainly come from
the determination of the inefficiencies in the readout and the knowledge  of the refractive index of the radiator.

\subsubsection {Analytical estimate of the yield}
The expected number of detected photo-electrons per unit length and per
unit of wavelength from a saturated track is given by:
\begin{multline}
\frac{d^2N}{d\lambda dx}  =  \frac{2 \pi \alpha}{\lambda^{2}} (1-
\frac{1}{n^{2}(\lambda)})\cdot {\rm QE_{\rm MaPMT}(\lambda)} \cdot
\hat{T}_{\rm lens}(\lambda) \\  
  \times \epsilon_{\rm geo} \cdot \epsilon_{\rm
  Mirror-refl}(\lambda) \cdot \epsilon_{\rm interface}(\lambda) \cdot \epsilon_{\rm
  threshold} \cdot \epsilon_{\rm pixel}
\end{multline}
where $\alpha\simeq \frac{1}{137}$ is the fine structure
constant, $n$ is the refractive index of the lens and
\begin{itemize}
\item ${\rm QE_{\rm MaPMT}(\lambda)}$: quantum efficiency of the MaPMTs measured in the laboratory (see Figure~\ref{fig:sr}).
\item $\hat{T}_{\rm lens}(\lambda)$: average transmission of the borosilicate lens measured in the laboratory and evaluated for the average photon path length.
\item $\epsilon_{geo}$: geometrical acceptance of the MaPMT with respect to the full ring, given by optical simulations.
\item $\epsilon_{\rm Mirror-refl} =0.9$: assumed coefficient of the reflective layer on the back surface of the lens. In first approximation it is considered independent of the wavelength. 
\item $ \epsilon_{\rm interface}$: transmission coefficient at the plane surface of the lens $\sim 96\%$, assumed independent of the wavelength
\item $\epsilon_{\rm threshold}$: average of the digital read out efficiency (see~\ref{subsec:threshold})
\item $\epsilon_{\rm pixel}$: dead area around each pixel (13\%).  
\end{itemize}
Integrating over the wavelength spectrum and the radiator length (13
mm) we obtain the expected number of photo-electrons. These are shown in Table~\ref{tab:PhotonYield}. There is a reasonable agreement
between the number of hits and the number expected. 
\begin{table}
  \centering
  \begin{tabular}{|c|c c|c c|c|}
    \hline
              & \multicolumn{2}{c|}{Data} &
              \multicolumn{2}{c|}{Simulation} &
              Analytical estimate \\\hline
              & mean & RMS & mean  & RMS & mean \\ \hline
      Total & 13.4 & 3.8     & 13.1 & 2.9 & - \\\hline
      PMT A   & 3.6 & 1.7   & 3.1  & 1.5 & 3.8 \\\hline
      PMT B & 3.3 & 1.6  & 3.1  & 1.5 & 4.1\\\hline
      PMT C  & 4.4 & 1.9 & 3.3  & 1.5 & 3.9\\\hline
      PMT D     & 4.2 & 1.8  & 3.3  & 1.5  & 4.1 \\\hline
     \hline
  \end{tabular}
   \caption{Comparison of average yield in each MaPMT from data,
     simulation and analytical estimate. For the
     analytical estimate the uncertainty on the transmission curve gives an error of $\pm 0.01$ photo-electrons on the yields. Other sources of error are negligible.}
  \label{tab:PhotonYield}
\end{table}
\subsubsection{Multi-track correlation studies}
\label{sec:telescopedata}
\begin{figure}[h]
\centering
\includegraphics[width=0.95\textwidth]{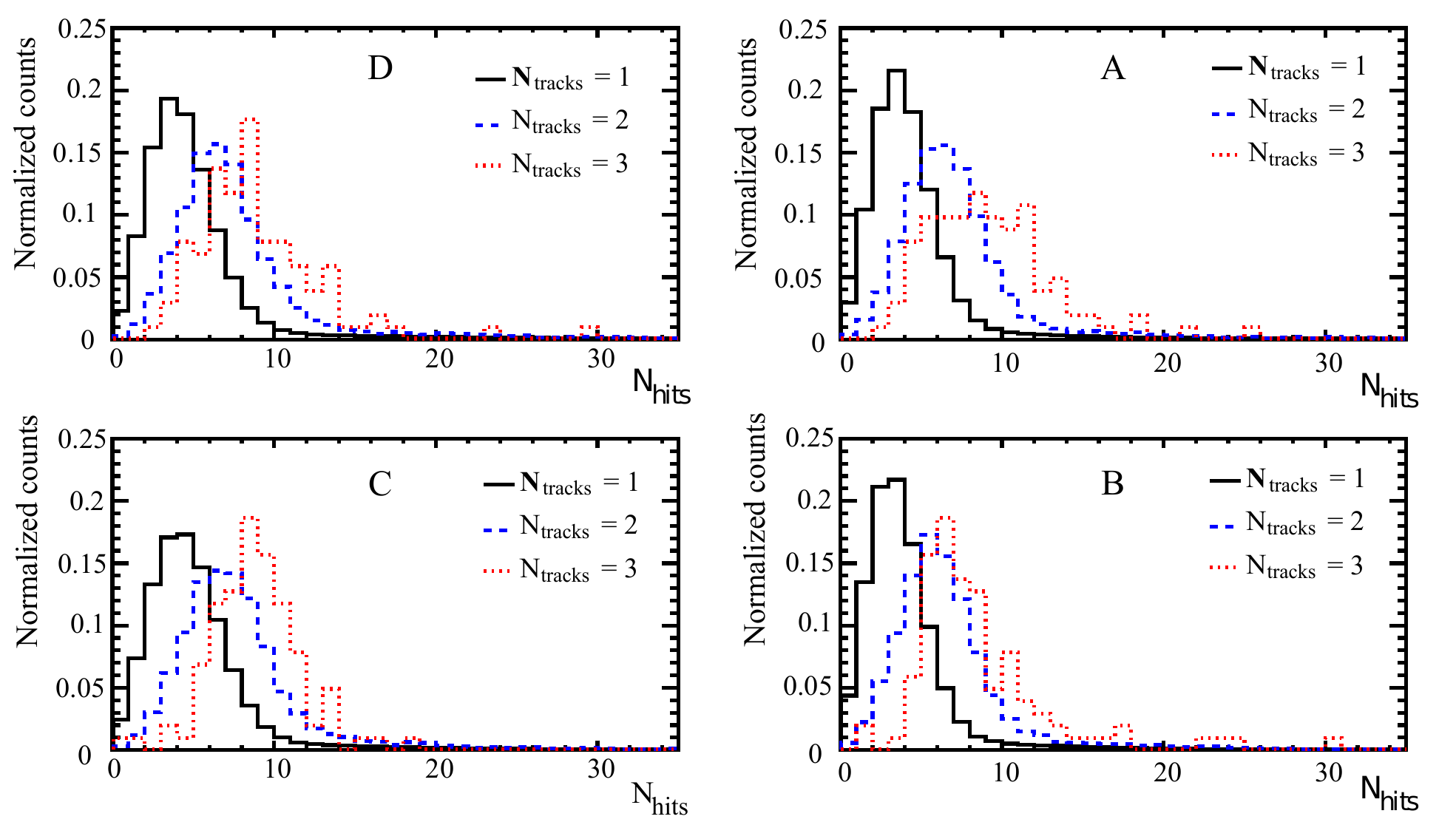}
\caption{Distributions of the number of hits in each MaPMT for events with one, two and three tracks associated.}
\label{fig:tel_nhitscorr2}
\end{figure}

A study of the correlation between multi-track events and number of
photo-electron hits on the MaPMTs has been
performed. Figure~\ref{fig:tel_nhitscorr2} shows the distribution of
the number of hits for each MaPMT for events that contained one, two
and three beam particle tracks. Table~\ref{tab:tel_nhitscorr2} shows the parameters of these distributions.
It is clear that as the number of tracks in the event increases so does the number of hits,
as expected. The number of photons is not proportional to the number of tracks
as each pixel is read out in binary mode and the probability that two photons will hit
the same pixel increases with the number of tracks. The correlation is in good agreement with the
simulation when the binary readout effect is included.

\begin{table}
\centering
\begin{tabular}{| c | c | c | c | c | c | c |}
\hline
 &   \multicolumn{2}{c|}{N$_{\text{hits, 1 track}}$} & \multicolumn{2}{c|}{N$_{\text{hits, 2 tracks}}$} & \multicolumn{2}{c|}{N$_{\text{hits, 3 tracks}}$}   \\
\hline
PMT & Mean & RMS  & Mean & RMS & Mean & RMS \\
\hline 
PMT A & 3.93 & 3.31 & 6.78 & 4.17 & 8.71 & 3.90\\
\hline
PMT B & 3.58 & 3.20 & 6.31 & 4.06 & 8.17 & 4.53\\
\hline
PMT C & 4.70 & 3.68 & 7.35 & 4.29 & 8.34 & 2.80\\ 
\hline
PMT D & 4.50 & 3.89 & 7.26 & 4.70 & 8.47 & 3.98\\ 
\hline
\end{tabular}
\caption{Parameters of the distributions of the number of hits for events with one, two and three tracks associated.}
\label{tab:tel_nhitscorr2}
\end{table}

\subsection{Fit of the ring}
\label{sec:fitring}
\begin{figure}[htbp]
\centering
\includegraphics[scale=0.4]{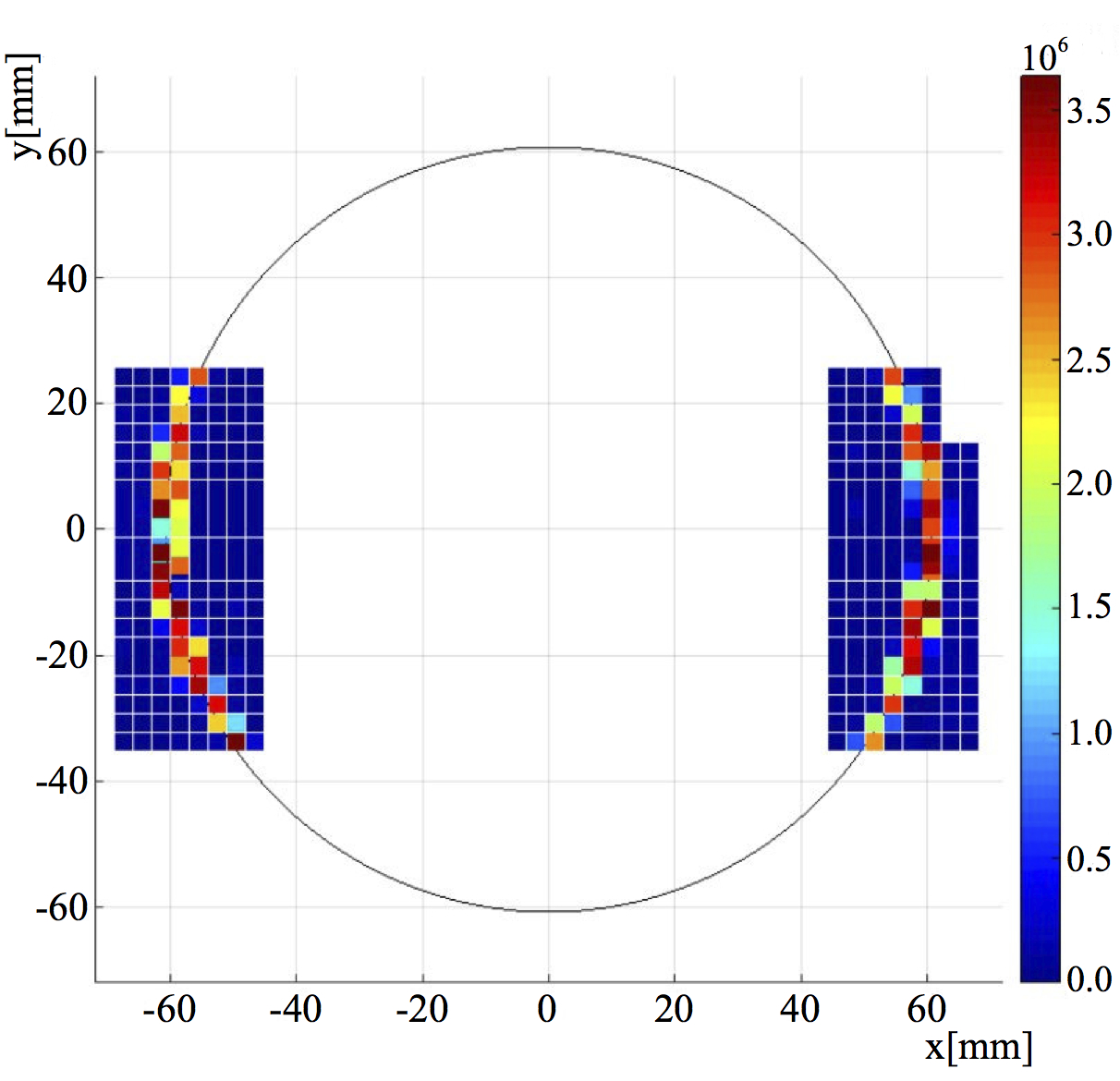}
\caption{The fitted ring superimposed to the integrated events. The
  figure shows the superimposition of two different runs taken with
  the ECs shifted in the vertical direction in order to illuminate
  different pixels of the MaPMTs.}
\label{fig:fitring}
\end{figure}

A ring fit procedure has been developed in order to obtain a robust value for the Cherenkov ring radius, centre and resolution and to
compare the data with the values expected from simulations.
Two different procedures have been developed. 
The first one uses the
integrated events. For each run, a minimization of the circle-to-data
distance weighted by the number of hits per pixel is performed on the accumulated number of hits per
pixel assuming a uniform distribution inside the pixel. The fitted
ring superimposed to the real data is shown in
Figure~\ref{fig:fitring} for two different runs. A radius of
$R=60.5\unitm{mm}$ and an RMS of 0.5\unitm{mm}, compatible with the values expected from simulations,
are obtained. 

\begin{figure}[htbp]
\centering
\includegraphics[scale=0.4]{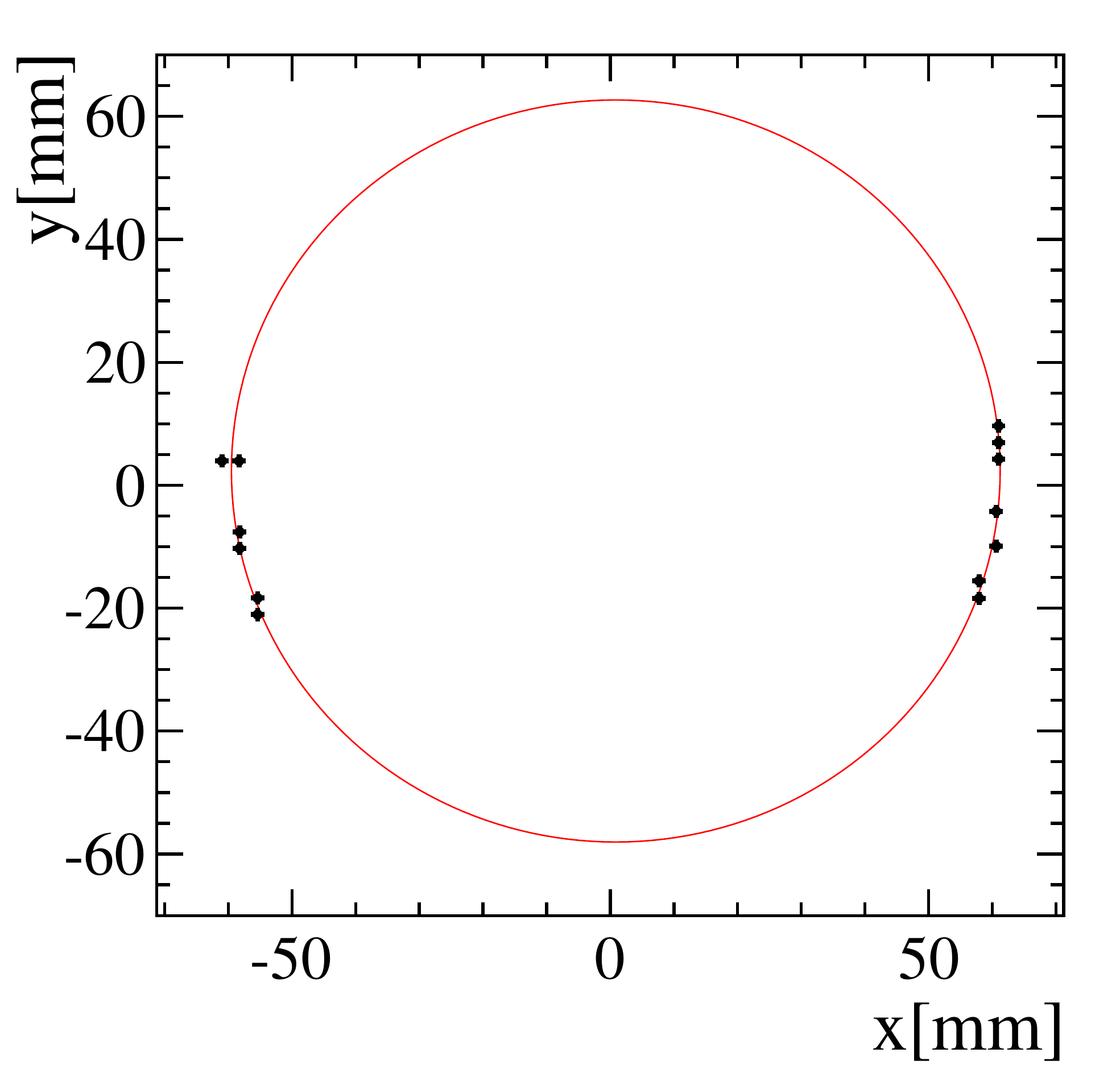}
\caption{Plot of the hits of a single event with the fitted ring
  superimposed.}
\label{fig:fitring1}
\end{figure}

\begin{figure}[htbp]
\centering
\includegraphics[scale=0.38]{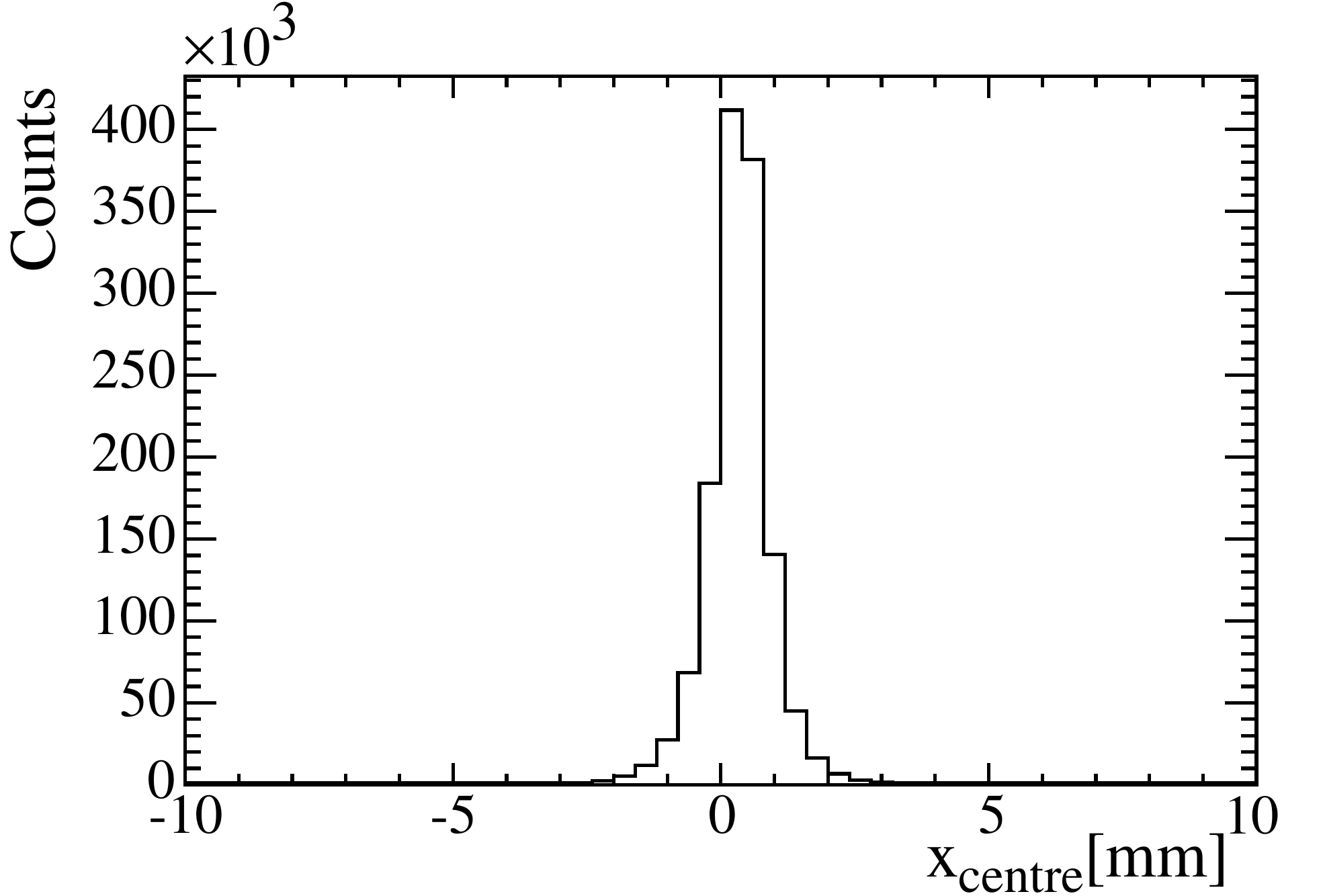}
\includegraphics[scale=0.38]{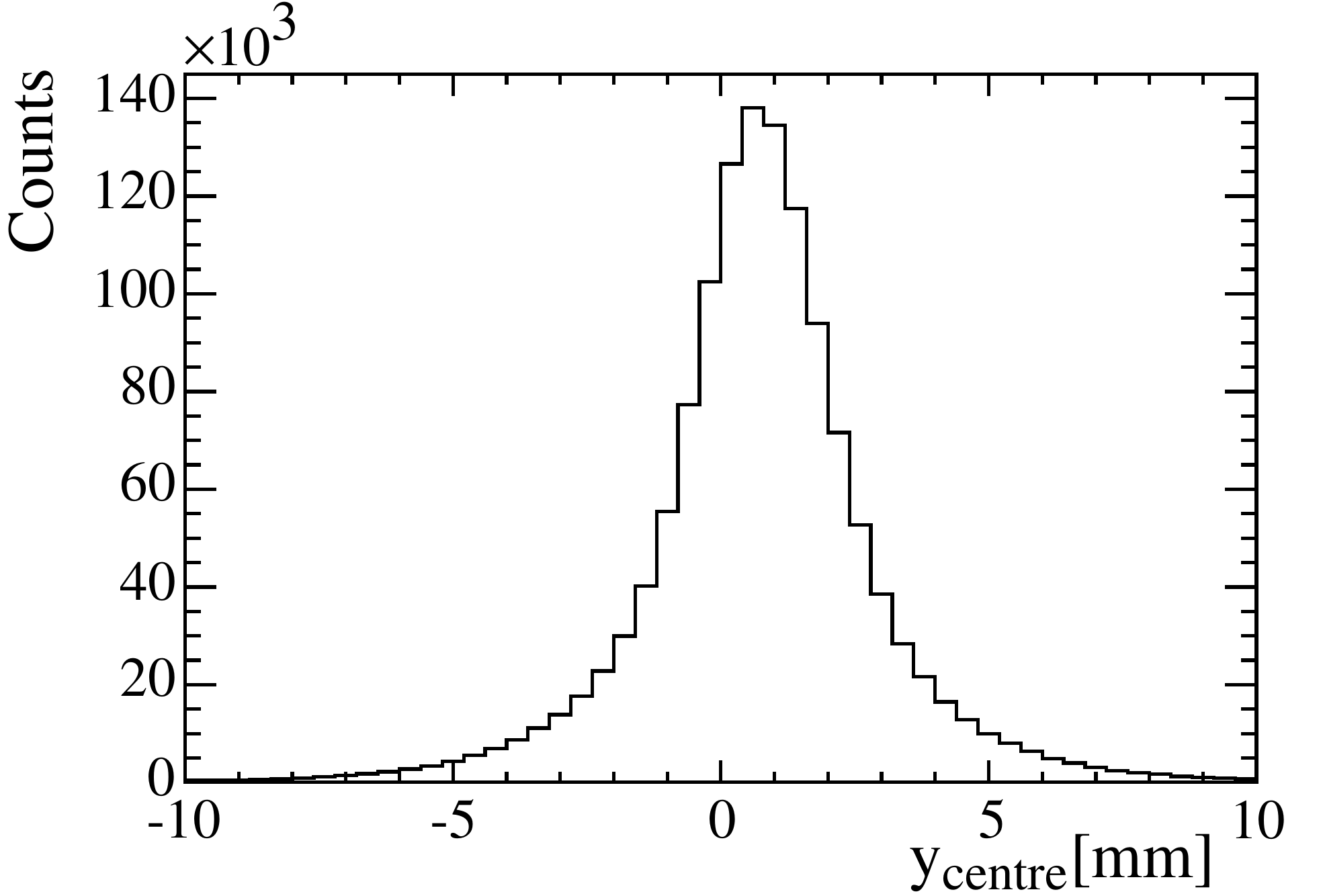}
\includegraphics[scale=0.38]{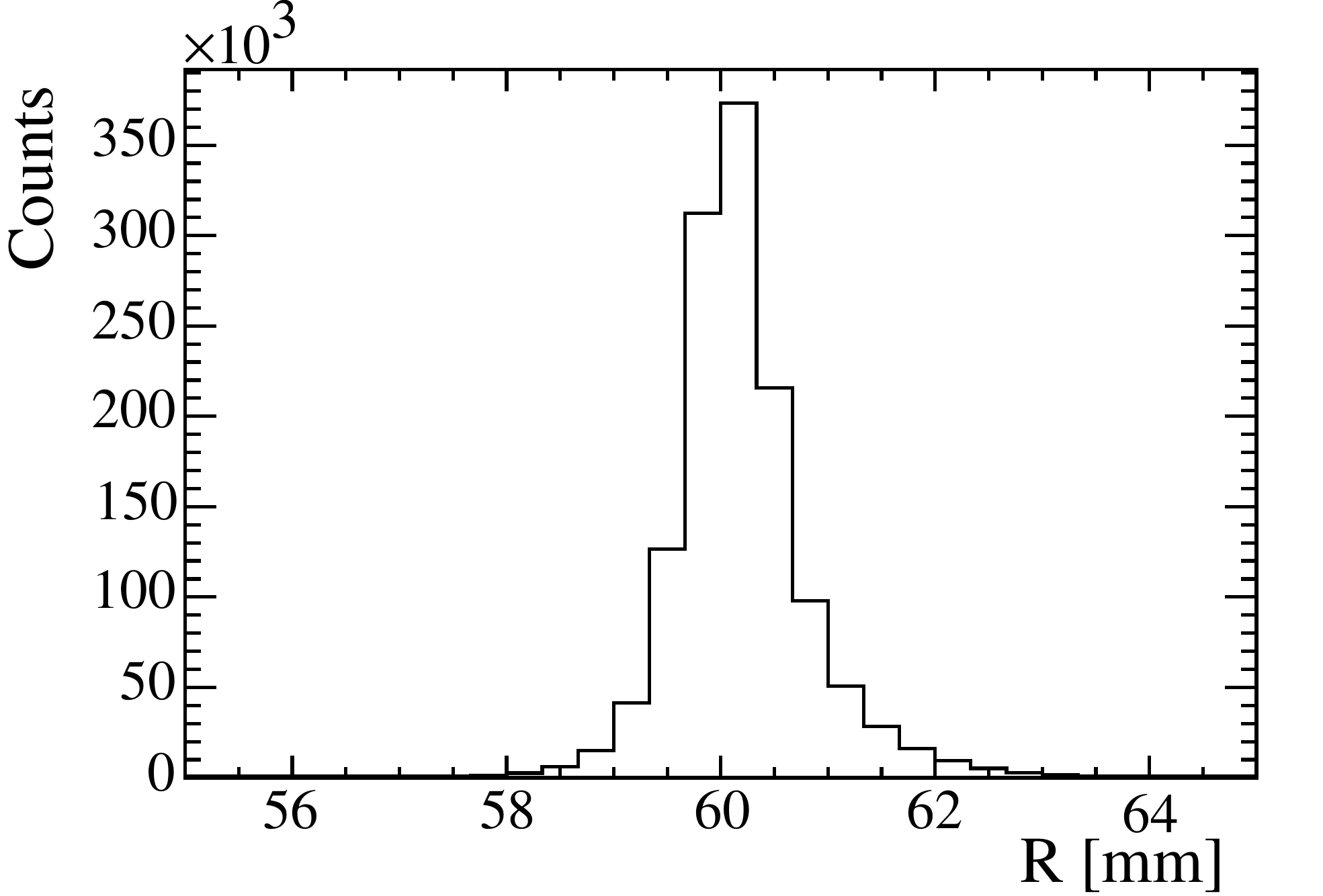}
\caption{Centre coordinates $(x,y)$ and radius distributions of
  reconstructed Cherenkov rings.}
\label{fig:xyr}
\end{figure}

The second procedure is based on fits to single events: each event is
fitted independently. The ring centre coordinates $(x,y)$
and the radius are the free parameters of the fit. The hit position is taken as the centre of the pixel with an error given by the pixel size. For each event the
distance between pixel centre and ring is minimised. A fit for a single event is
shown in Figure~\ref{fig:fitring1}.
Distributions of the centre coordinates and of the radius
extracted from the single event fits are reported in
Figure~\ref{fig:xyr}.
The mean value of the radius of $60.3\unitm{mm}$ and the
sigma of the distribution $\sigma_{\rm R} = 0.5 \unitm{mm}$ are in 
good agreement with the values calculated with the first method and 
with the values expected from the simulation.

\subsection{Cherenkov angle reconstruction}
\label{sec:cherenkovangle}

The reconstruction of the Cherenkov angle is performed using the algorithm described in~\cite{Roger:1999}, which requires as inputs the hit detection point, 
the photon emission point, the centre of curvature of the mirror and the direction of the charged track. In this case the `mirror' is the curved surface of the lens where
there is a reflective coating and its radius of curvature is the same as that of the lens. The algorithm makes use of the fact that one can define a 
plane of reflection for the photons reflected at the mirror using the
three input coordinates and that the mirror reflection point is also
in this plane defined by these
three coordinates.

The photon emission point is taken to be the mid-point of the track
segment radiating detectable photons. To simplify the geometry in the
reconstruction, the photon's reflection on the flat surface of the
lens is accounted for by using an (optically equivalent) virtual emission point outside of the lens. The virtual emission point is the reflection of the emission point in the flat surface of the lens.
The photons undergo a refraction when exiting the radiator towards the MaPMT plane. 
An ``image plane'' is considered where the photons would have traveled the same optical path length, 
without this refraction.  Using simulated data, every pixel centre on the MaPMT plane is mapped to this ``image plane''. The hits mapped on to 
this plane are used as the detection point for the reconstruction  algorithm.  These transformations for the emission point and detection point 
ensure that the three input coordinates are in the
plane as required by the algorithm described in~\cite{Roger:1999}.

\subsubsection {Results from reconstruction }

\begin{figure}[h!!!]
\centering
\includegraphics[scale=0.5]{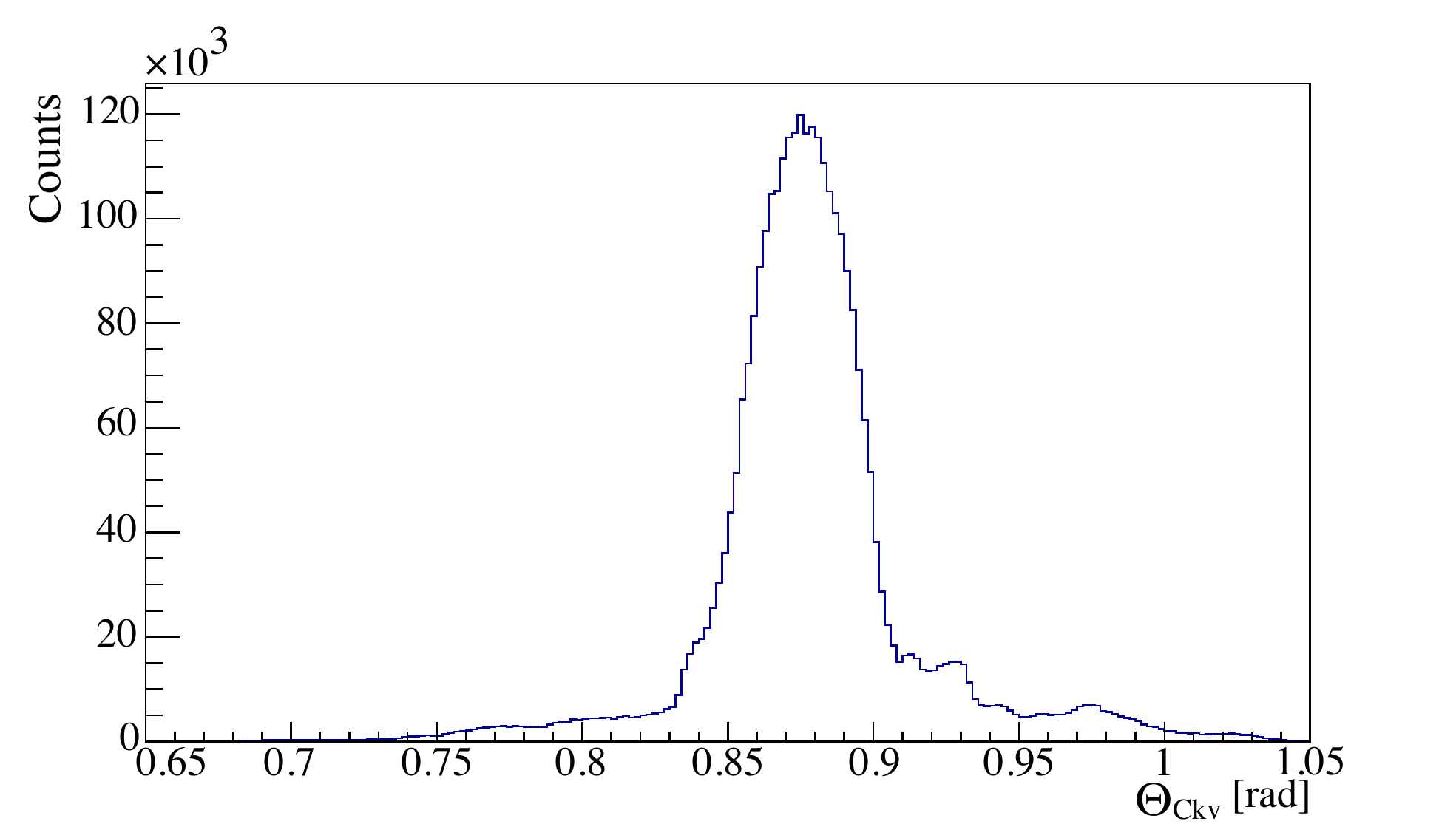}
\caption{Reconstructed Cherenkov angle from real data. This plot contains the data from all four MaPMTs. A Gaussian fit to the main peak has a mean of 875 mrad and a width of 17 mrad. 
  }
\label{fig:RealData-recon-ckv}
\end{figure}

\begin{figure}[h!!!]
\centering
\includegraphics[scale=0.5]{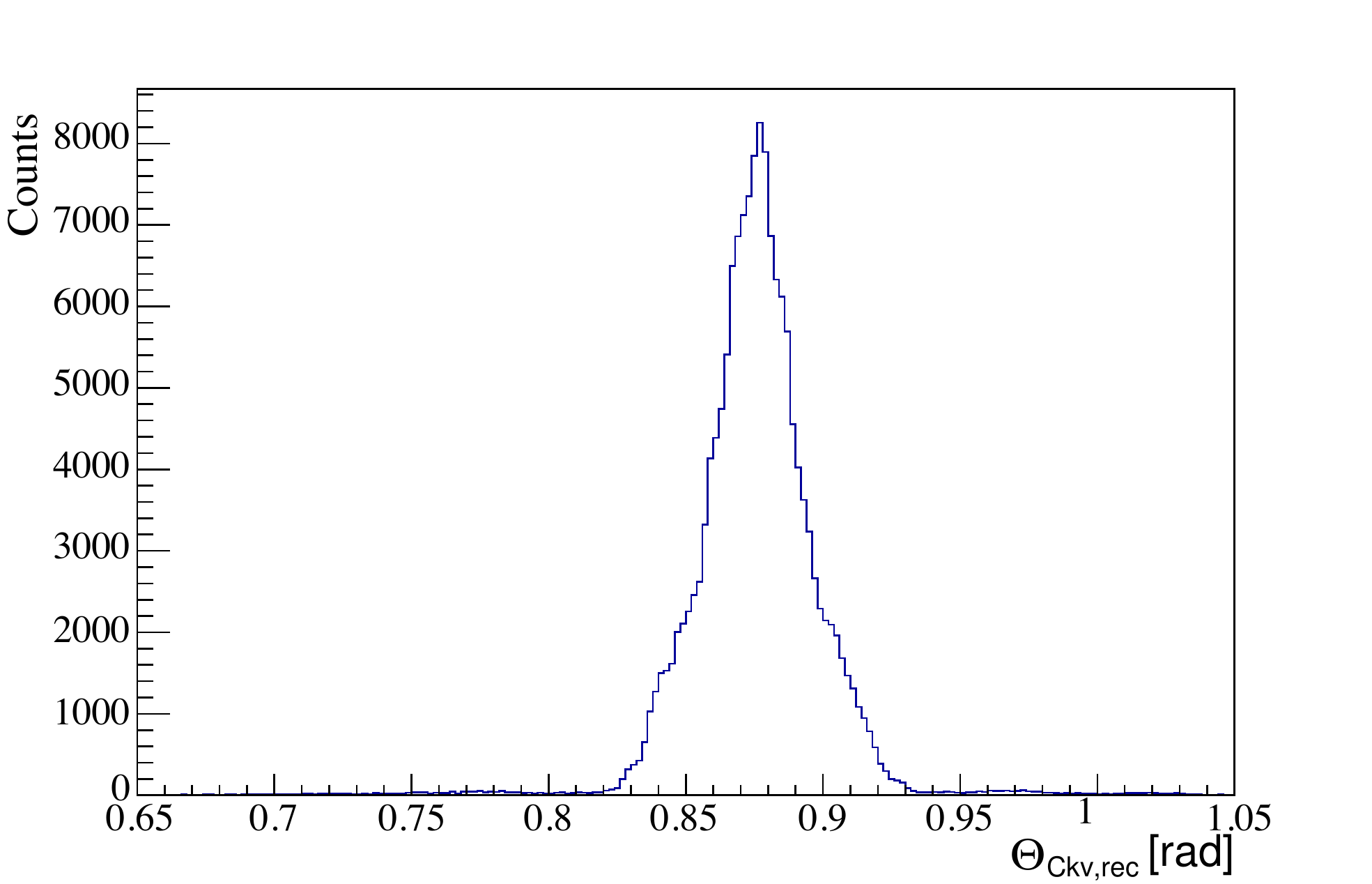}
\caption{Reconstructed Cherenkov angle from full simulation. This plot contains the data from all four MaPMTs. A Gaussian fit to the main peak has a mean of 875 mrad and a width of 17 mrad.  }
\label{fig:MC-recon-ckv}
\end{figure}

In Figure~\ref{fig:RealData-recon-ckv} the distribution of the reconstructed Cherenkov angle is shown for data 
and in Figure~\ref{fig:MC-recon-ckv}  the same quantity is shown for
the simulated data. The two distributions have been fitted in the range
0.84 -- 0.91~mrad with a Gaussian function. The values of the fitted mean and width are quoted in the corresponding captions. From these figures it can be seen that resolution in data is compatible with that from simulation. 
The components of the resolution from simulated data are shown in Table~\ref{tab:CkvRes}, which indicate that the 
resolution is dominated by the pixel size. The chromatic error comes
from the variation of refractive index of the radiator with
wavelength.
The pixel size effect also results in the small structures seen 
in the tail background regions in Figure~\ref{fig:RealData-recon-ckv} and Figure~\ref{fig:MC-recon-ckv}.

\begin{table}
  \centering
  \begin{tabular}{|c|c|}
    \hline
               & Resolution \\
               &  mrad        \\\hline  
      Chromatic & 4.5  \\\hline
      Emission point  & 7.4   \\\hline
      Pixel size  & 15.8  \\\hline
      Total simulation    & 17.0 $ \pm$ 1.3  \\\hline \hline
      Data &  17.0 $\pm $ 0.8 \\\hline
    
  \end{tabular}
   \caption{Cherenkov angle resolutions from data and simulations. The components contributing to the resolution 
     are also listed. The overall resolution and its uncertainty are dominated by the pixel size contribution.}
  \label{tab:CkvRes}
\end{table}

\clearpage
\clearpage

\clearpage
\section{Conclusions}
\label{sec:conclusions}
The results of the beam tests at the CERN SPS 
using the baseline photon detectors
and electronics for the LHCb RICH Upgrade have been presented. Using
a lens as a solid Cherenkov radiator at the same time as a focusing
element, Cherenkov rings have been observed. The setup has been used
to test the main concept for the integration of the electronics and
photon detectors in a complete readout system able to take data
at a readout rate of 40~MHz, including the compact design of
four MaPMTs into an Elementary Cell.
The opto-electronics chain was similar to that proposed for the photon detector array of the RICH upgrade.
Results show the capability to detect single photons, with a low dark
current. The measured cross-talk level is low and compatible with the
RICH upgrade requirements.
The spatial resolution obtained from the
reconstruction of the Cherenkov ring showed a good agreement with simulations. Further characterisation of new
prototypes will be carried out in future beam tests,
progressively integrating more modular units and with improved Data Acquision and Detector Control Systems.
\clearpage

\section*{Acknowledgements}
\noindent 
The authors would like to thank all the LHCb technical staff in the universities and institutes and especially at CERN who contributed to the RICH test-beam activity. We gratefully acknowledge the CERN SPS machine operators for the delivery of stable and reliable beams. The Science Technology and Facilities Council (STFC) and Istituto Nazionale di Fisica Nucleare (INFN) are also acknowledged for funding this work.\\We also thank T.~Gys, K.~Wyllie, H.~Schindler, M.~van Beuzelkom, T.~Schneider, C.~David and F.~Fontanelli.\\This project has also received funding from the European Union's Horizon
2020 Research and Innovation programme under Grant Agreement no. 654168.

\clearpage
\addcontentsline{toc}{section}{References}
\setboolean{inbibliography}{true}
\bibliographystyle{LHCb}
\bibliography{main,LHCb-DP,LHCb-TDR}

\ifx\mcitethebibliography\mciteundefinedmacro
\PackageError{LHCb.bst}{mciteplus.sty has not been loaded}
{This bibstyle requires the use of the mciteplus package.}\fi
\providecommand{\href}[2]{#2}
\begin{mcitethebibliography}{1}
\mciteSetBstSublistMode{n}
\mciteSetBstMaxWidthForm{subitem}{\alph{mcitesubitemcount})}
\mciteSetBstSublistLabelBeginEnd{\mcitemaxwidthsubitemform\space}
{\relax}{\relax}

\bibitem{Alves:2008zz}
LHCb collaboration, A.~A. Alves~Jr.\ {\em et~al.},
  \ifthenelse{\boolean{articletitles}}{\emph{{The \lhcb detector at the LHC}},
  }{}\href{http://dx.doi.org/10.1088/1748-0221/3/08/S08005}{JINST \textbf{3}
  (2008) S08005}\relax
\mciteBstWouldAddEndPuncttrue
\mciteSetBstMidEndSepPunct{\mcitedefaultmidpunct}
{\mcitedefaultendpunct}{\mcitedefaultseppunct}\relax
\EndOfBibitem
\bibitem{LHCb-DP-2012-003}
M.~Adinolfi {\em et~al.},
  \ifthenelse{\boolean{articletitles}}{\emph{{Performance of the \lhcb RICH
  detector at the LHC}},
  }{}\href{http://dx.doi.org/10.1140/epjc/s10052-013-2431-9}{Eur.\ Phys.\ J.\
  \textbf{C73} (2013) 2431}, \href{http://arxiv.org/abs/1211.6759}{{\tt
  arXiv:1211.6759}}\relax
\mciteBstWouldAddEndPuncttrue
\mciteSetBstMidEndSepPunct{\mcitedefaultmidpunct}
{\mcitedefaultendpunct}{\mcitedefaultseppunct}\relax
\EndOfBibitem
\bibitem{LHCb-TDR-014}
LHCb collaboration, \ifthenelse{\boolean{articletitles}}{\emph{{LHCb PID
  Upgrade Technical Design Report}}, }{}
  \href{http://cdsweb.cern.ch/search?p=CERN-LHCC-2013-022&f=reportnumber&actio%
n_search=Search&c=LHCb+Reports} {CERN-LHCC-2013-022}.
\newblock LHCb-TDR-014\relax
\mciteBstWouldAddEndPuncttrue
\mciteSetBstMidEndSepPunct{\mcitedefaultmidpunct}
{\mcitedefaultendpunct}{\mcitedefaultseppunct}\relax
\EndOfBibitem
\bibitem{Cadamuro:2014hza}
L.~Cadamuro {\em et~al.},
  \ifthenelse{\boolean{articletitles}}{\emph{{Characterization of the Hamamatsu
  R11265-103-M64 multi-anode photomultiplier tube}},
  }{}\href{http://dx.doi.org/10.1088/1748-0221/9/06/P06021}{JINST \textbf{9}
  (2014) P06021}, \href{http://arxiv.org/abs/1403.3215}{{\tt
  arXiv:1403.3215}}\relax
\mciteBstWouldAddEndPuncttrue
\mciteSetBstMidEndSepPunct{\mcitedefaultmidpunct}
{\mcitedefaultendpunct}{\mcitedefaultseppunct}\relax
\EndOfBibitem
\bibitem{Carniti:2012ue}
P.~Carniti {\em et~al.},
  \ifthenelse{\boolean{articletitles}}{\emph{{CLARO-CMOS, a very low power ASIC
  for fast photon counting with pixellated photodetectors}},
  }{}\href{http://dx.doi.org/10.1088/1748-0221/7/11/P11026}{JINST \textbf{7}
  (2012) P11026}, \href{http://arxiv.org/abs/1209.0409}{{\tt
  arXiv:1209.0409}}\relax
\mciteBstWouldAddEndPuncttrue
\mciteSetBstMidEndSepPunct{\mcitedefaultmidpunct}
{\mcitedefaultendpunct}{\mcitedefaultseppunct}\relax
\EndOfBibitem
\bibitem{Agostinelli:2002hh}
Geant4 collaboration, S.~Agostinelli {\em et~al.},
  \ifthenelse{\boolean{articletitles}}{\emph{{Geant4: a simulation toolkit}},
  }{}\href{http://dx.doi.org/10.1016/S0168-9002(03)01368-8}{Nucl.\ Instrum.\
  Meth.\  \textbf{A506} (2003) 250}\relax
\mciteBstWouldAddEndPuncttrue
\mciteSetBstMidEndSepPunct{\mcitedefaultmidpunct}
{\mcitedefaultendpunct}{\mcitedefaultseppunct}\relax
\EndOfBibitem
\bibitem{Roger:1999}
LHCb collaboration, R.~Forty, \ifthenelse{\boolean{articletitles}}{\emph{{RICH
  pattern recognition for LHCb}}, }{}Nucl.\ Instrum.\ Meth.\  \textbf{A433}
  (1999) 257\relax
\mciteBstWouldAddEndPuncttrue
\mciteSetBstMidEndSepPunct{\mcitedefaultmidpunct}
{\mcitedefaultendpunct}{\mcitedefaultseppunct}\relax
\EndOfBibitem
\end{mcitethebibliography}

\newpage

\end{document}